\documentclass[12pt]{article}
\pdfoutput=1

\usepackage[T1]{fontenc} 
\usepackage{microtype} 

\usepackage[hmarginratio=1:1,top=32mm,columnsep=20pt]{geometry} 
\usepackage{fancyhdr} 
\usepackage{abstract}

\usepackage{titlesec} 
\titleformat{\section}[block]{\large\bfseries\centering}{\thesection}{1em}{} 
\titleformat{\subsection}[block]{\bfseries}{\thesubsection}{1em}{} 

\usepackage{color}
\definecolor{dark-gray}{gray}{0.20}
\definecolor{gray}{gray}{0.30}
\definecolor{light-gray}{gray}{0.80}
\definecolor{dark-red}{rgb}{0.7,0,0}
\definecolor{dark-green}{rgb}{0.1,0.4,0}
\definecolor{dark-blue}{rgb}{0.3,0.3,0.7}
\definecolor{light-blue}{rgb}{0.8,0.8,1}

\usepackage{cite}
\usepackage{hyperref}
\hypersetup{
	colorlinks=true,
	linkcolor=dark-blue,
	citecolor=dark-red,
	urlcolor=dark-green,
	linktoc=page
}

\usepackage{enumerate}
\usepackage{graphicx}
\usepackage[percent]{overpic}
\usepackage{tikz}
\usepackage{subcaption}
\usepackage{skak}
\usepackage{staves}

\usepackage{amsmath,amssymb,slashed}
\usepackage{amsthm}

\numberwithin{equation}{section}
\usepackage{slashed}

\newcommand{\dd}{\mathrm{d}}

\newcommand{\e}{\mathrm{e}}

\newcommand{\rme}{\mathrm{e}}
\newcommand{\rmd}{\mathrm{d}}

\newcommand{\be}{\begin{equation}}
\newcommand{\ee}{\end{equation}}
\newcommand{\bea}{\begin{eqnarray*}}
\newcommand{\eea}{\end{eqnarray*}}

\newcommand{\f}[2]{\frac{#1}{#2}}

\newcommand{\SU}{\mathop{\rm SU}}
\newcommand{\SO}{\mathop{\rm SO}}

\newcommand{\U}{\mathop{\rm {}U}}

\title{\vspace{-0mm}\fontsize{20pt}{24pt}\selectfont\textbf{Holographic Uniformization\\ and Black Hole Attractors}\vspace{3mm}}

\author{
\large{
\href{mailto:nikolay.bobev@kuleuven.be}{Nikolay Bobev$^{\text{\runictext{a}}}$,} \href{mailto:ffg@kuleuven.be}{Fri{\dh}rik Freyr Gautason$^{\text{\runictext{a}},\text{\runictext{b}}}$,} and \href{mailto:k.parmentier@columbia.edu}{Klaas Parmentier$^{\text{\runictext{a}},\text{\runictext{c}}}$}}\\[6mm]
\normalsize $^\text{\runictext{a}}$Instituut voor Theoretische Fysica, KU Leuven\\
\normalsize Celestijnenlaan 200D, 3001 Leuven, Belgium\\[3mm]
\normalsize $^\text{\runictext{b}}$University of Iceland, Science Institute\\
\normalsize Dunhaga 3, 107 Reykjav{\'i}k, Iceland\\[3mm]
\normalsize $^\text{\runictext{c}}$Department of Physics, Columbia University\\
\normalsize 538 West 120th Street, New York, NY 10027, USA\\[3mm]
}

\date{}

\begin{document}  
{\hypersetup{urlcolor=black}\maketitle}
\thispagestyle{empty}

\begin{abstract}
\noindent We establish an attractor mechanism for the horizon metric of asymptotically locally AdS$_4$ supersymmetric black holes. The K\"ahler moduli of the metric are not fixed in the asymptotic region, however supersymmetry dictates that the near horizon metric must be the constant curvature one. We show how this mechanism is realized in detail for four-dimensional $\mathcal{N}=2$ gauged supergravity coupled to vector multiplets by focusing on the STU model. A similar analysis is performed for gauged supergravity theories in five, six, and seven dimensions where we establish the same mechanism by extending previous results on holographic uniformization.

\end{abstract}

\newpage

\setcounter{tocdepth}{2}
\tableofcontents

 \section{Introduction}
 \label{sec:introduction}
  
The dynamics of the branes in string and M-theory at low energy is captured by the physics of supersymmetric quantum field theories. An alternative vantage point on the same physics is offered by the supergravity solutions obtained by considering a large number of branes and studying their near horizon limit. This dichotomy is the basis of the AdS/CFT correspondence and its many generalizations and applications. A prominent set of explicit examples of AdS/CFT, which is the main focus of this work, arises from the dynamics of branes wrapped on compact manifolds \cite{Maldacena:2000mw,Bershadsky:1995qy}, see \cite{Gauntlett:2003di,Naka:2002jz} for a review and a number of explicit solutions.  
 
Our interest here is in situations where the branes are wrapped on a smooth compact Riemann surface $\Sigma_{\mathfrak{g}}$. In order to preserve supersymmetry, the QFT on the world-volume of the branes has to be partially topologically twisted. As reviewed in \cite{Bobev:2017uzs} this topological twist leads to an RG flow across dimensions interpolating between the $(d+1)$-dimensional QFT in the UV and a $(d-1)$-dimensional QFT in the IR. Here we consider situations in which both the UV and the IR QFTs are superconformal. The supergravity description of this setup is captured by a domain wall solution which interpolates between an asymptotically locally AdS$_{d+2}$ region in the UV and an AdS$_{d}$ near horizon region in the IR. This holographic RG flow solution can also be viewed as a black brane background in $d+2$ dimensions with a horizon topology determined by $\Sigma_{\mathfrak{g}}$.

Whenever the partial topological twist of an SCFT with a continuous R-symmetry is well-defined, it can be performed for an arbitrary metric on the Riemann surface while still preserving the same amount of supersymmetry~\cite{Witten:1988ze}. When interpreted holographically this immediately implies that the black brane solutions should exist for any choice of metric on $\Sigma_{\mathfrak{g}}$. Indeed, this was shown to be the case in \cite{Anderson:2011cz} where several examples of such holographic RG flows arising from wrapped D3- and M5-branes were studied in detail. The main result in \cite{Anderson:2011cz} is that while the metric on $\Sigma_{\mathfrak{g}}$ can be arbitrary in the asymptotically locally AdS$_{d+2}$ UV region, the supergravity BPS equations act as a geometric flow which uniformizes the metric on the Riemann surface. In other words only the constant curvature metric is allowed in the near horizon AdS$_{d}$ region.

This uniformization behavior of the holographic RG flows across dimensions in \cite{Anderson:2011cz} confirms, in a non-trivial manner, two important physical expectations. First, it was conjectured in \cite{Gaiotto:2009we}, see also \cite{Gaiotto:2011xs}, that the K\"ahler moduli of the Riemann surface should appear as irrelevant deformations of the $(d-1)$-dimensional IR CFT. This conjecture is explicitly confirmed for the examples analyzed in \cite{Anderson:2011cz}. This in turn is compatible with the more general expectation that the number of degrees of freedom in a QFT should decrease along an RG flow from the UV to the IR. Second, the black brane solutions have non-trivial entropy and general arguments from black hole thermodynamics dictate that this entropy should not depend on continuous parameters. This is akin to the situation in asymptotically flat black holes in string theory for which the attractor mechanism, see \cite{Ferrara:1995ih,Strominger:1996kf,Ferrara:1996dd} and \cite{Sen:2007qy} for a nice review, ensures that this conundrum does not arise. Therefore, the solutions studied in \cite{Anderson:2011cz} can also be interpreted as an explicit realization of an attractor mechanism for the moduli associated with the horizon geometry of asymptotically AdS black branes.

Our goal in this paper is to generalize and extend the results of \cite{Anderson:2011cz} to other wrapped brane solutions in string and M-theory and thus establish a myriad of novel examples of holographic uniformization. We study numerous new holographic RG flow solutions arising from M2-, D2-, D4-, and M5-branes wrapped on $\Sigma_{\mathfrak{g}}$. We perform these studies in appropriate consistent truncations of ten- and eleven-dimensional supergravity to a gauged supergravity theory in lower dimensions. In all cases we analyze, we find that the supergravity BPS equations admit arbitrary smooth metrics on $\Sigma_{\mathfrak{g}}$ in the asymptotically AdS$_{d+2}$ UV region. On the other hand, in the near horizon AdS$_{d}$ IR region the only regular solutions are the ones with a constant curvature metric on the Riemann surface. This behavior is precisely the same as the holographic uniformization, or attractor mechanism, for the metric on $\Sigma_{\mathfrak{g}}$ expected from the results in \cite{Anderson:2011cz}.

We separate the technical analysis of the supergravity BPS equations into two main parts. For the minimal gauged supergravity solutions in four, five, six, and seven dimensions the BPS equations reduce to a single non-linear PDE for a function of three variables which determines the metric and matter fields in the full supergravity background. One can then analyze this single PDE and arrive at the uniformization behavior described above. This is very similar to the results in \cite{Anderson:2011cz}. We also study solutions of matter coupled supergravity theories in four, five, and seven dimensions. In these examples the BPS equations are considerably more involved and take the form of a system of coupled PDEs for functions of three variables. Nevertheless we are able to analyze these BPS equations in the UV and IR regions of the geometry and establish the expected uniformization behavior in all cases we studied.

Our results in four-dimensional gauged supergravity deserve some additional comments. In this case we find asymptotically locally AdS$_4$ supersymmetric static black holes with a $\Sigma_{\mathfrak{g}}$ horizon. The freedom to choose an arbitrary metric on $\Sigma_{\mathfrak{g}}$ at asymptotic infinity can be interpreted as black hole ``hair'', see \cite{Bekenstein:1996pn} for a review of the ``no-hair'' conjectures. The possibility to have black hole ``hair'' in asymptotically AdS$_4$ solutions is well-known, in particular in the context of holographic superconductors  \cite{Gubser:2008px,Hartnoll:2008vx}. However, our solutions are distinct from holographic superconductors since we do not employ charged scalar fields to construct them. The attractor mechanism for asymptotically AdS$_4$ black holes with scalar fields has been discussed extensively in the literature, starting with \cite{Cacciatori:2009iz,DallAgata:2010ejj,Hristov:2010ri}. We note that these results have a very different character than the attractor mechanism for asymptotically flat black holes. The reason is that in the asymptotically AdS$_4$ region, the values of the scalar fields are typically fixed to specific values determined by the structure of the potential of the gauged supergravity theory. Therefore, these scalars do not represent continuous parameters on which the AdS$_4$ black hole entropy can depend. In contrast, the metric moduli associated to $\Sigma_{\mathfrak{g}}$ we study here are continuous and therefore holographic uniformization can be viewed as a more direct analog of the attractor mechanism for asymptotically flat black holes. Finally, we note that one can use supersymmetric localization results to account for the black hole entropy of large classes of static supersymmetric asymptotically locally AdS$_4$ black holes, see \cite{Benini:2015eyy,Benini:2016rke,Azzurli:2017kxo} and \cite{Zaffaroni:2019dhb} for a recent review. These results were established by studying black hole solutions with constant curvature metric on $\Sigma_{\mathfrak{g}}$. Our more general black hole solutions have the same area of the horizon as the solutions in \cite{Benini:2015eyy,Azzurli:2017kxo} and therefore carry the same entropy. This in turn is compatible with the fact that the metric on $\Sigma_{\mathfrak{g}}$ is a $Q$-exact deformation of the topologically twisted index used to account for the black hole entropy \cite{Benini:2015noa,Benini:2016hjo,Closset:2016arn}.

We organize the presentation of our results as follows. In the next section we explain in some detail our main results by focusing on the so called universal black holes as defined in \cite{Bobev:2017uzs}. In Section~\ref{sec:STU} we extend these results by coupling the supergravity theory to additional matter fields. We discuss a number of generalizations and open problems that stem from our work in Section~\ref{sec:conclusions}. In the Appendix we present in some detail the derivation of the supergravity BPS equations for the universal black holes discussed in Section~\ref{sec:universal}.

 \section{Universal black holes }
 \label{sec:universal}

The simplest examples of the asymptotically locally AdS$_{d+2}$ black hole and black brane solutions we are interested in are the so-called universal RG flows across dimensions studied in \cite{Bobev:2017uzs}. The supergravity solutions realize a partial topological twist of a $(d+1)$-dimensional SCFT with a continuous R-symmetry placed on the manifold ${\bf R}^{d-1}\times \Sigma_g$. In this construction some supersymmetry is preserved by cancelling the curvature of the $\U(1)_\Sigma$ structure group of the Riemann surface $\Sigma_g$  by a background gauge field for the $\U(1)_R$ subgroup of the R-symmetry. Spinor parameters that are singlet with respect to the twisted group $\text{diag}[\U(1)_\Sigma\times\U(1)_R]$ lead to preserved supersymmetry \cite{Witten:1988ze}. As shown in \cite{Maldacena:2000mw} these partial topological twists can be realized holographically by asymptotically locally AdS$_{d+2}$ supergravity solutions with an ${\bf R}^{d-1}\times \Sigma_g$ boundary and a non-trivial magnetic flux for the dynamical  gauge field dual to the Abelian R-current. 

In this section we will present such supergravity solutions in four, five, six and seven dimensions. To illustrate our construction we focus on the minimal gauged supergravity theories which allow for this type of solutions. These theories contain the dynamical bulk fields in the gravity multiplet which are dual to the energy momentum multiplet of the dual SCFT. We study more general supergravity theories coupled to matter multiplets in Section~\ref{sec:STU}. An important difference between our analysis and the discussion in \cite{Bobev:2017uzs} is that we allow for the metric on the Riemann surface to be arbitrary. A general metric on the Riemann surface is compatible with the topological twist and does not break additional supersymmetry and is therefore natural to consider in supergravity. Indeed, this was explored in a holographic context in \cite{Anderson:2011cz} and our analysis and results bear resemblance to \cite{Anderson:2011cz}. We have assumed that the metric on the Riemann surface is smooth throughout our calculations, see \cite{Bobev:2019ore} for a recent discussion of similar supergravity solutions that have Riemann surfaces with point-like singularities.

While the details of our construction depend on the dimension in which the supergravity theory lives there are notable similarities in the derivation of the BPS equations that follow from the supersymmetry variations of the theory. We present the details of the derivation of these BPS equations in Appendix~\ref{UniversalDerivation}. Here we note that the supersymmetric solutions we study have supersymmetry transformation parameter that obey projectors of the following schematic form
\be
\gamma_{\hat{x}\hat{y}}\epsilon=\Gamma\epsilon\,,\quad \gamma_{\hat{\rho}} \epsilon = \epsilon\,.
\ee
Here $\gamma_{\hat\mu}$ are space-time gamma matrices, $x,y$ denote local coordinates on the Riemann surface, $\rho$ is the radial coordinate of AdS, and $\Gamma$ is a matrix that acts on the internal indices of the spinor $\epsilon$ associated with the R-symmetry.

We now proceed with the analysis of the BPS equations and their solutions for each of these four minimal gauged supergravities.

 \subsection{Four dimensions}
 \label{4DUniversal}

We start our exploration with the minimal ${\cal N}=2$ gauged supergravity theory in four dimensions \cite{Freedman:1976aw,Fradkin:1976xz}. The bosonic field content consists of the metric and a Maxwell field $A$. We present the bosonic Lagrangian of the theory in \eqref{eq:4dminLagapp}. As described above we look for static supersymmetric solutions which realize the partial topological twist in the dual SCFT. The derivation of the BPS equations is carried out in Appendix \ref{4Dapp} and here we only display the results. The complete supersymmetric solution can be written in terms of a single function, $\varphi$, which depends on the radial coordinate $\rho$ as well as the Riemann surface coordinates $x$ and $y$. The metric and the gauge field of these supersymmetric solutions take the compact form 
\be\label{4Dsolution}
\begin{split}
\dd s_4^2 =&\, -\f{\dd t^2}{(\partial_\rho\varphi)^2} + (\partial_\rho\varphi)^2\big(\dd\rho^2 + \e^{4\varphi}(\dd x^2 + \dd y^2)\big)\,,\\
A =&\, (\partial_x\varphi) \dd y - (\partial_y\varphi) \dd x\,,
\end{split}
\ee
where $\varphi$ satisfies the non-linear partial differential equation
\be\label{4Dmaster}
\triangle \varphi +\e^{4\varphi}\big(\partial_\rho^2 \varphi + (\partial_\rho \varphi)^2\big)=0\,,
\ee
and we have defined $\triangle \equiv \partial^2_x + \partial^2_y$.

\subsubsection{The constant curvature black hole}
\label{subsubsec:cc4dBH}

An exact analytic solution of the equation in \eqref{4Dmaster} is given by the black hole solution of \cite{caldarelli:1999}, see also \cite{Romans:1991nq,Gauntlett:2001qs}, which has a constant curvature metric on the Riemann surface. To obtain this solution we write the function $\varphi$ as a sum of $\rho$-dependent function, $\mathfrak{f}(\rho)$,  and a function on the Riemann surface $g(x,y)$
\be\label{4DroundAnsatz}
\varphi = \f12\big( \mathfrak{f}(\rho) + g(x,y)\big)\,.
\ee
The function $g(x,y)$ defines a metric on the Riemann surface 
\be\label{RiemannSurfaceMetric}
\dd s_{\Sigma_\mathfrak{g}}^2 = \e^{2g}(\dd x^2 + \dd y^2)\,.
\ee
Using the separable Ansatz \eqref{4DroundAnsatz} in \eqref{4Dmaster} yields the following differential equations 
\be\label{4DLiouville}
\triangle g + \kappa\,\e^{2g} =0 \,,\qquad \e^{2\mathfrak{f}}\big(2\mathfrak{f}''+(\mathfrak{f}')^2\big)=2\kappa\,,
\ee
where the prime denotes a derivative with respect to $\rho$. The constant $\kappa$ is initially introduced to separate the equation \eqref{4Dmaster} but has a simple geometric interpretation. The sign of $\kappa$ determines the curvature of the Riemann surface in \eqref{RiemannSurfaceMetric}, indeed the Ricci scalar of the metric $\dd s_{\Sigma_\mathfrak{g}}^2$ is simply $2\kappa$. Shifting the function $\mathfrak{f}$ by a constant it is possible to normalize $\kappa$ such that it takes the values $\kappa = -1,0,1$. In this paper we assume that the Riemann surface is compact and smooth which means that if the genus $\mathfrak{g}=0$ then $\kappa =1$, if $\mathfrak{g}=1$ then $\kappa=0$, and if $\mathfrak{g}\ge 2$ then $\kappa =-1$. More explicitly the solutions of the first equation in \eqref{4DLiouville} for the constant curvature metric on the covering space of the Riemann surface can be written as
\begin{equation}\label{eq:CCSigmag}
\begin{split}
g(x,y) &= -\log y \,, \qquad\qquad\qquad\qquad\qquad \mathfrak{g}\geq 2\,,\\
g(x,y) &= {\rm const}\,, \qquad\qquad\qquad\qquad\qquad~~ \mathfrak{g}=1 \,,\\
g(x,y) &= -\log (1+x^2+y^2) +\log 2\,, \qquad \mathfrak{g}=0\,.
\end{split}
\end{equation}
The solution of the second equation in \eqref{4DLiouville}, up to shifts in the radial coordinate $\rho$ and rescaling of the time coordinate, is
\be\label{4Dfroundsol}
\e^{\mathfrak{f}} = \rho^2 + \f{\kappa}{2}\,.
\ee
With this explicit solution for $\varphi$ in \eqref{4DroundAnsatz} we find the following metric and gauge field
\be\label{4DBHsol}
\begin{split}
\dd s_4^2 &= -\bigg(\rho + \f{\kappa}{2\rho}\bigg)^2\dd t^2 + \bigg(\rho + \f{\kappa}{2\rho}\bigg)^{-2}\dd\rho^2 + \rho^2 \dd s_{\Sigma_g}^2\,, \\
F &= -\frac{\kappa}{2} \rme^{2g} \dd x\wedge \dd y\,.
\end{split}
\ee
For $\rho\to\infty$ we have an asymptotically locally AdS$_4$ spacetime with ${\bf R}^{1,1}\times \Sigma_{\mathfrak{g}}$ boundary. As $\rho$ decreases we encounter a naked singularity for $\kappa \ge 0$ at $\rho=0$ \cite{Romans:1991nq}. Since we are interested in regular black hole solutions we take $\kappa = -1$ which has an AdS$_2$ near horizon region located at $\rho^2 = 1/2$ around which the function $\mathfrak{f}$ takes the form
\be
\e^{\mathfrak{f}} = \sqrt{2}\rho - 1 + {\cal O}( \sqrt{2}\rho - 1)^2\,.
\ee
The near horizon metric then reads
\be\label{4DIR}
\dd s_4^2 = \f14 \dd s_{\text{AdS}_2}^2 + \f12 \dd s_{\Sigma_{\mathfrak{g}}}^2\,,
\ee
where  both two-dimensional metrics $\dd s_{\text{AdS}_2}^2$ and $\dd s_{\Sigma_{\mathfrak{g}}}^2$ are normalized such that their Ricci scalar equals $-2$. We refer to this supersymmetric black hole solution with $\kappa=-1$ and $\mathfrak{f}$ given in \eqref{4Dfroundsol} as the constant curvature black hole. Borrowing terminology from holography we will refer to the asymptotically AdS$_4$ region as ``the UV region'' and the near horizon AdS$_2$ region as ``the IR region''. The black hole in \eqref{4DBHsol} has a finite Bekenstein-Hawking entropy which, in the semiclassical approximation, can be accounted for microscopically by embedding it in string or M-theory and employing holography and supersymmetric localization \cite{Azzurli:2017kxo}. Note also that it follows from \eqref{4DLiouville} that complex structure deformations of the metric on $\Sigma_{\mathfrak{g}}$ leave the solution invariant and do not affect the horizon area and thus the black hole entropy.

Our next goal is to analyze small perturbations around the constant curvature black hole solution that satisfy the equation \eqref{4Dmaster}. 

\subsubsection{Perturbative analysis}
\label{subsubsec:4Dperturbations}
We consider linearized perturbations around the solution \eqref{4DroundAnsatz} and \eqref{4Dfroundsol}. We write 
\be
\varphi = \f12 \Big(\log(\rho^2-1/2) + g(x,y)\Big) + \delta \varphi\,,\quad \triangle g - \e^{2g} =0\,,
\ee
where we have chosen $\kappa=-1$ and $\delta\varphi(\rho,x,y)$ represents a small fluctuation. Inserting this expression into \eqref{4Dmaster} and expanding to linear order in $\delta\varphi$, we obtain the partial differential equation
\be\label{4Dperturbations}
\Big[(\e^{-2g}\triangle-2)+2\rho(\rho^2-1/2)\partial_\rho+(\rho^2-1/2)^2\partial_\rho^2\Big]\delta\varphi =0\,.
\ee
To solve this differential equation it is useful to define the operator
\begin{equation}
\triangle_g=\e^{-2g}\triangle\,,
\end{equation}
which is the Laplacian on the Riemann surface $\Sigma_{\mathfrak{g}}$ with metric \eqref{RiemannSurfaceMetric}. Since \eqref{4Dperturbations} is linear, it is useful to  decompose the fluctuations into eigenmodes of the Laplacian as follows: 
\begin{equation}\label{modeExpansion}
   \delta{ \varphi}(\rho,x,y) =\sum^\infty_{n=0} {\varphi}_n(\rho)Y_n(x,y)\,, \quad \triangle_g Y_n = -\mu_n Y_n, \quad\mu_n\ge 0\,.
\end{equation}
Note that since we have a smooth and compact metric on the Riemann surface the eigenvalues of the Laplacian $\mu_n$ are non-negative.\footnote{There may be stronger lower bounds on the eigenvalues $\mu_n$, see \cite{Camporesi:1994ga}, which will not be important for our analysis.} 
Using orthogonality of the eigenfunctions $Y_n$ the equations for $\varphi_n(\rho)$ take the form
\be
\Big[-(\mu_n+2)+2\rho(\rho^2-1/2)\partial_\rho+(\rho^2-1/2)^2\partial_\rho^2\Big]\varphi_n(\rho) =0\,.
\ee
This equation admits the following analytic solution which depends on two sets of integration constants
\be\label{4Dperturbedphiaandb}
\varphi_n(\rho) = a_n \Big(\f{\sqrt{2}\rho-1}{\sqrt{2}\rho+1}\Big)^{\gamma_n}+ b_n \Big(\f{\sqrt{2}\rho-1}{\sqrt{2}\rho+1}\Big)^{-\gamma_n}\,,\quad \gamma_n=\sqrt{1+\f{\mu_n}{2}}\,.
\ee
Notice that $\gamma_n >0$ and since the deformation must be regular at the horizon $\sqrt{2}\rho \to 1$ we should choose the integration constants $b_n=0$. The linearized perturbation around the constant curvature black hole solution is therefore 
\be\label{4Dperturbedphi}
\delta{ \varphi} =\sum^\infty_{n=0} a_n \Big(\f{\sqrt{2}\rho-1}{\sqrt{2}\rho+1}\Big)^{\gamma_n} Y_n(x,y)\,,
\ee
where $a_n$ are undetermined real constants which should be small in order to ensure that validity of the linearized approximation. In general the constants $\gamma_n$ are irrational numbers and one may worry whether the corresponding solutions have curvature singularities. We have checked explicitly that this is not the case. 

\begin{figure}
\centering
\begin{overpic}[width=0.3\textwidth]{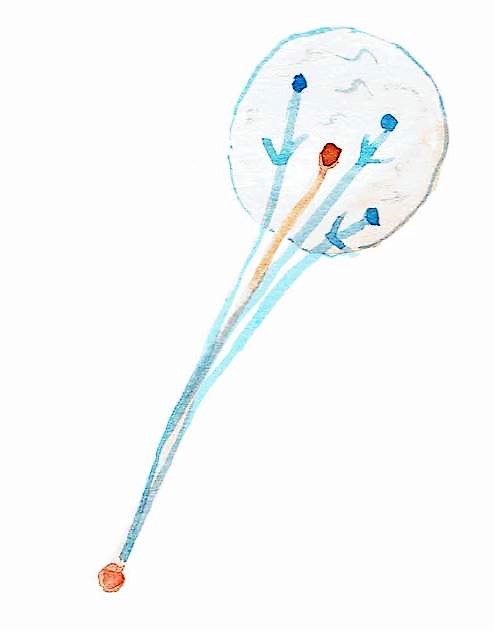}
\put (5,5) {IR}
\put (27,85) {UV}
\end{overpic}
\caption{\label{4Dflow}The uniformizing nature of the equation \eqref{4Dmaster} demonstrated in a cartoon. A generic point on the moduli space of solutions represents a choice of metric on the Riemann surface which is not constant curvature in the UV region. In the IR all these solutions approach the constant curvature curvature black hole solution and thus the metric on the Riemann surface is uniformized.}
\end{figure}

The perturbative solution in \eqref{4Dperturbedphi} demonstrates clearly the general behavior the BPS solutions we study in this work. In the UV region, $\rho\to\infty$, the perturbations are completely unconstrained and are controlled by the constants $a_n$. The choice of constants $a_n$ represents a choice of a metric on the Riemann surface $\Sigma_{\mathfrak{g}}$ in the UV. Instead of the constant curvature metric $g$ that obeys \eqref{4DLiouville}, the UV metric is determined by
\be\label{eq:gUV4d}
g_\text{UV} = g + 2 \sum^\infty_{n=0} a_nY_n(x,y)\,.
\ee
Clearly this function does not satisfy the Liouville equation \eqref{4DLiouville} and is therefore not a constant curvature metric. The metric in \eqref{eq:gUV4d} represents a single point on the \emph{moduli space} of metrics we can choose on the Riemann surface. As $\rho$ decreases and we approach the IR region, the perturbations in \eqref{4Dperturbedphi} get smaller and smaller and ultimately vanish near the horizon at $\rho\to 1/\sqrt{2}$. In this sense the metric on the Riemann surface is uniformized by the PDE in \eqref{4Dmaster} as we approach the IR region. This qualitative behavior is illustrated in Figure \ref{4Dflow}. Here we demonstrated this uniformizing behavior using  a linearized approximation around the constant curvature black hole solution. Nevertheless, based on the results in \cite{Anderson:2011cz}, we expect the same result to hold for arbitrary perturbations of the UV metric.  To support this expectation we now perform a general UV expansion of the equation \eqref{4Dmaster}. 

\subsubsection{UV analysis}
\label{subsubsec:4dUV}

Another simple exact solution of the PDE in \eqref{4Dmaster} with metric as in \eqref{4Dsolution} is provided by the AdS$_4$ vacuum given by
\be
\varphi = \log\rho\,.
\ee
The black hole solution in \eqref{4DBHsol} is asymptotically locally AdS$_4$ which is manifested by the following UV expansion of the function $\varphi$ as $\rho \to\infty$
\be
\varphi = \log\rho + \f12 g(x,y) + {\cal O}(1/\rho^2)\,,
\ee
where $g(x,y)$ is the constant curvature metric on the Riemann surface. We now demonstrate that a general solution of \eqref{4Dmaster} has exactly the same type of UV expansion where the function $g$ is \emph{not} constrained to satisfy the Liouville equation in \eqref{4DLiouville}. To avoid confusion we will denote this more general metric by $\hat{g}(x,y)$. In other words, any smooth metric is permitted on the Riemann surface in the UV. In the UV region a general solution of \eqref{4Dmaster} can be written as as the following expansion
\be\label{eq:4dUVexp}
\varphi =\log\rho + \f12 \hat{g}(x,y) + \f{1}{\rho}v(x,y) + \sum_{n\ge 2} \rho^{-n}g_n(x,y)\,.
\ee
Solving \eqref{4Dmaster} order by order in $\rho$ as $\rho\to\infty$ we find, that $\hat{g}(x,y)$ as well as $v(x,y)$ are unconstrained. However, all other functions $g_n(x,y)$ are all algebraically determined in terms of (derivatives of) $\hat{g}(x,y)$ and $v(x,y)$. For instance, the first two functions take the form
\be
\begin{split}
g_2 =& -\f12v^2 -\f14 \triangle_{\hat{g}} \hat{g}\,,\\
g_3 =& \f13 v^3 + \f12 v\triangle_{\hat{g}} \hat{g} - \f16 \triangle_{\hat{g}} v\,.
\end{split}
\ee
The existence of two free functions in this UV expansion is compatible with general expectations for asymptotically locally AdS$_4$ solutions that follow from the Fefferman-Graham expansion \cite{FG,Anderson:2004yi}. These two functions can also be interpreted in the dual field theory. The function $\hat{g}(x,y)$ deforms the metric on the Riemann surface and thus serves as a source for the energy momentum tensor in the dual SCFT, while the function $v$ controls the vev of this operator which in turn determines the state of the IR QFT. For generic choice of functions $\hat{g}$ and $v$ we expect the full solution of \eqref{4Dmaster} to be singular in the IR region, this intuition is based on the expected general behavior of gravitational solutions as well as previous results on holographic RG flows \cite{Gubser:2000nd}. However, if the vev function $v(x,y)$ is carefully chosen, the flow solution of the PDE in \eqref{4Dmaster} will reach the unique regular AdS$_2\times \Sigma_{\mathfrak{g}}$ solution, with metric \eqref{4DIR}, in the IR region. This result is compatible with intuition from the dual QFT where it is expected that to arrive at conformal dynamics for the IR theory one has to carefully tune the vev of the relevant operators triggering the RG flow in the UV.  Let us see how this pans out for the perturbative solution in \eqref{4Dperturbedphi}. Expanding this solution in the UV region we find
\be\label{eq:UVexppert4d}
\varphi  =\log\rho + \f12 g(x,y)+\sum^\infty_{n=0} a_nY_n(x,y) - \f{1}{\rho} \sum^\infty_{n=0} a_n\sqrt{2+\mu_n}Y_n(x,y) + {\cal O}(1/\rho^2)\,.
\ee
This explicitly demonstrates that for supergravity solutions which approach the AdS$_2\times \Sigma_{\mathfrak{g}}$ solution in the IR region the function $v$ in \eqref{eq:4dUVexp} is precisely determined by the metric on the Riemann surface, given by the first two terms on the right-hand-side of \eqref{eq:UVexppert4d}. If we change the function of $(x,y)$ in the $\rho^{-1}$ term in \eqref{eq:UVexppert4d} the resulting solution will be singular in the IR. This singular behaviour should be captured by keeping both the $a_n$ and $b_n$ coefficients in the perturbative solution \eqref{4Dperturbedphiaandb}. 

We can summarized our results as follows. We have an initial value problem in the UV region for the PDE in \eqref{4Dmaster} which is controlled by two arbitrary functions of the Riemann surface, $\hat{g}(x,y)$ and $v(x,y)$, in \eqref{eq:4dUVexp}. The solutions of this PDE that lead to regular supergravity backgrounds are such that in the IR the metric on the Riemann surface is uniformized to the constant curvature one. To be more precise we can formulate this holographic uniformization principle as follows: \emph{given any smooth metric on the Riemann surface, $\hat{g}(x,y)$, there exists a unique function $v(x,y)$ such that the solution to the BPS equation \eqref{4Dmaster} with $\hat{g}$ and $v$ as initial values is regular and approaches the near horizon metric in \eqref{4DIR}.}  An alternative formulation of the same statement is that the constant curvature metric of the black hole horizon in \eqref{4DIR} is an attractor in the moduli space of metrics for all supersymmetric static black holes with regular horizons. We emphasize the need to tune the subleading term in the Fefferman-Graham expansion in order to obtain a regular IR solution is by no means a special feature of our supergravity solutions. This is a general property of holographic RG flows which is perhaps more familiar in the context discussed in \cite{Gubser:2000nd}.

As explained around \eqref{4Dperturbedphi}, we have established a perturbative proof of this uniformization principle. A non-perturbative global proof should be constructed using methods similar to the ones employed in \cite{Anderson:2011cz}. After illustrating the general behavior of the uniformization flows we are interested in we move on to discuss supersymmetric black brane solutions in higher dimensions which exhibit similar structure.

 \subsection{Five dimensions}
 \label{subsec:5dmin}

In this section we repeat the analysis above for supersymmetric static solutions of the minimal ${\cal N}=2$ supergravity in five dimensions \cite{DAuria:1981yvr,Chamseddine:1980sp}. The bosonic content of the theory is the same as in four dimensions, see Appendix~\ref{5Dapp} for the explicit form of the bosonic Lagrangian. The analysis of the BPS equations and their solutions is very similar to the one above and thus we keep the discussion brief.  

The derivation of the BPS solution is carried out in Appendix \ref{5Dapp}. This results in the following solution for the bosonic fields 
\be
\begin{split}
\dd s_5^2  =&\, \f{\dd s_{\mathbf{R}^{1,1}}^2}{(\partial_\rho\varphi)} + (\partial_\rho\varphi)^2\big(\dd\rho^2 + \e^{6\varphi}(\dd x^2 + \dd y^2)\big)\,,\\
A =&\, (\partial_x\varphi) \dd y - (\partial_y\varphi) \dd x\,,
\end{split}
\ee
where $\varphi$ satisfies the partial differential equation
\be\label{5Dmaster}
\triangle \varphi +\e^{6\varphi}\big(\partial_\rho^2 \varphi + 2(\partial_\rho \varphi)^2\big)=0\,.
\ee
Notice the similarity with \eqref{4Dmaster} which only differs from \eqref{5Dmaster} by numerical factors.

\subsubsection{The constant curvature black string}

We start our analysis by studying the analytic black string solution found in \cite{klemm2000,Maldacena:2000mw}, see also \cite{Benini:2013cda,Benini:2015bwz}. This solution is again obtained by assuming that $\varphi$ can be written as a sum
\be\label{5DroundAnsatz}
\varphi = \f13\big( \mathfrak{f}(\rho) + g(x,y)\big)\,.
\ee
The function $g(x,y)$ defines a metric on the Riemann surface as in \eqref{RiemannSurfaceMetric}.
Assuming that the equation \eqref{5Dmaster} is separable leads
\be\label{5DLiouville}
\triangle g + \kappa\,\e^{2g} =0 \,,\qquad \e^{2\mathfrak{f}}\big(2\mathfrak{f}''+3(\mathfrak{f}')^2\big)=3\kappa\,,
\ee
where $\kappa$ denotes the curvature of the Riemann surface and the explicit form of the metric on the covering space is given in \eqref{eq:CCSigmag}. The equation for $\mathfrak{f}$ can be solved explicitly by first changing coordinates as follows 
\be\label{5Dcoords}
r(\rho) \equiv \f{1}{3} \e^{\mathfrak{f}(\rho)} \mathfrak{f}'(\rho)\,,
\ee
which, using \eqref{5DLiouville}, implies
\be
\dd \rho = \f{3\e^{\mathfrak{f}}\dd r}{3r^2 +\kappa}\,,\qquad \partial_r \mathfrak{f} = \f{9r}{3r^2 +\kappa}\,.
\ee
The five-dimensional solution then takes the explicit form
\be\label{5DBSsol}
\begin{split}
\dd s_5^2 &= r^\frac12 \Big(r+\frac{\kappa}{3r}\Big)^\frac32 \dd s_{\mathbf{R}^{1,1}} + \Big(r +\frac{\kappa}{3r}\Big)^{-2}\dd r^2 + r^2 \dd s_{\Sigma_{\mathfrak{g}}}^2\,, \\
F &= -\frac{\kappa}{3} \rme^{2g} \dd x\wedge \dd y\,.
\end{split}
\ee
where we have absorbed the only integration constant into redefinitions of the coordinates on $\mathbf{R}^{1,1}$. This metric is singular for $\kappa = 0,1$ but has a smooth horizon given by a hyperbolic Riemann surface of constant curvature for $\kappa =-1$. As expected, for a supersymmetric black string solution the metric in the near horizon region, $r\to 3^{-1/2}$, has an AdS$_3$ factor and takes the form
\be\label{5DIR}
\dd s_5^2 = \f49 \dd s_{\text{AdS}_3}^2 + \f13 \dd s_{\Sigma_\mathfrak{g}}^2\,,
\ee
where both metrics are normalized such that $R_{ij} = -(d-1) g_{ij}$.

\subsubsection{Perturbative analysis}
\label{subsubsec:5Dperturbations}
We now study linearized perturbations around the solution \eqref{5DBSsol} using the equation \eqref{5Dmaster}. As in four dimensions it is convenient to expand the perturbations in eigenfunctions of the Laplacian on the Riemann surface as in \eqref{modeExpansion}. Our starting point is therefore
\be\label{5DperturbationsAnsatz}
\varphi = \f13\big( \mathfrak{f}(\rho) + g(x,y)\big) + \sum^\infty_{n=0} {\varphi}_n(\rho)Y_n(x,y)\,,
\ee
where $\mathfrak{f}$ and $g$ define the constant curvature black string and $Y_n$ are defined in \eqref{modeExpansion}. We assume that the functions $\varphi_n$ remain small for all values of $\rho$. Linearizing \eqref{5Dmaster} we then find the following differential equation for $\varphi_n$
\be
\Big[-(\mu_n+2)+\f13 \e^{2\mathfrak{f}}(4\mathfrak{f}' \partial_\rho+3\partial_\rho^2)\Big]\varphi_n(\rho) =0\,.
\ee
Using the coordinate transformation \eqref{5Dcoords}, we obtain the equation
\be
\Big[-(\mu_n+2)+\f19 (3r^2-1)(9 r \partial_r + (3r^2-1)\partial_r^2)\Big]\varphi_n(r) =0\,.
\ee
This equation has only one regular solution given by
\be\label{eq:phisolhyper}
\varphi_n = a_n (3r^2-1)^{\gamma_n}{}_2F_1\big[\gamma_n,\gamma_n+1; \tfrac32+2\gamma_n;1-3r^2 \big]\,,
\ee
where we have defined
\begin{equation}\label{eq:5dmingamman}
\gamma_n = \frac{\sqrt{25+12\mu_n}-1}{4}  
\end{equation}
and $a_n$ are arbitrary real constants. Note that since $\mu_n\geq0$ we find that the constants $\gamma_n$ are positive. In the IR region we have $r\to 3^{-1/2}$ and therefore from \eqref{eq:phisolhyper} we find
\be
\varphi_n \to  a_n (3r^2-1)^{\gamma_n} \to 0\,,
\ee
since $\gamma_n>0$. This shows that the perturbations vanish in the IR and the metric uniformizes to the constant curvature one in \eqref{5DIR}.

\subsubsection{UV analysis}
\label{subsubsec:5dminUV}

The UV analysis is very similar to the one performed in Section~\ref{subsubsec:4dUV} and leads to the same conclusions. In the UV region, $\rho \to \infty$, the general solution to \eqref{5Dmaster} takes the form 
\be\label{eq:5dUVphi}
\varphi =\f12 \log\rho + \f13 \hat{g}(x,y) + \frac{1}{\rho}v(x,y)+ \sum_{n\ge 1}\sum_{m=1}^{n-1} \rho^{-n}(\log\rho)^m g_{n,m}(x,y)\,.
\ee
We can then solve \eqref{5Dmaster} order by order for large $\rho$. This leads to relations between the function $g_{n,m}$ and the unconstrained functions $\hat{g}(x,y)$ and $v(x,y)$. All functions $g_{n,m}(,y)$ can be expressed algebraically in terms of (derivatives of) $\hat{g}(x,y)$ and $v(x,y)$. For example the lowest order function is
\be
g_{1,1} = \f13 \triangle_{\hat{g}} \hat{g}\,.
\ee
A notable difference between the expansion in \eqref{eq:5dUVphi} and the four-dimensional one in \eqref{eq:4dUVexp} is the presence of $\log\rho$ terms in \eqref{eq:5dUVphi}. These terms are characteristic for the Fefferman-Graham expansion for odd-dimensional asymptotically locally AdS spaces and their presence can be traced to the conformal anomaly in the dual quantum field theory.

This UV expansion leads to the same conclusion as in Section~\ref{subsubsec:4dUV}. Namely, we find that in the UV region the metric on the Riemann surface can be arbitrary and is not constrained to obey the Liouville equation. To find regular solutions one needs to adjust the function $v(x,y)$ appropriately and then one finds that in the IR region the solutions approaches the near horizon geometry of the supersymmetric black string in \eqref{5DIR} with a constant curvature metric on the Riemann surface. We thus conclude that the PDE in \eqref{5Dmaster} leads to the same uniformization behavior as discussed below  \eqref{eq:UVexppert4d}.

 \subsection{Six dimensions}
 \label{subsec:6dmin}

We now turn to supersymmetric black brane solutions of the six-dimensional minimal gauged supergravity constructed in \cite{Romans:1985tw}. The difference here with respect to the two previous examples is that the minimal supergravity in six dimensions has a larger gravity multiplet with bosonic content a scalar field in addition to the metric and an $\SU(2)$ gauge field.

We present the derivation of the BPS equations for this theory in Appendix \ref{6Dapp}. The result is the following configuration for the bosonic fields  
\be
\begin{split}
\dd s_6^2  =&\, \f{\e^{2\rho}\,\dd s_{\mathbf{R}^{1,2}}^2}{(\partial_\rho\varphi)^{1/2}} + (\partial_\rho\varphi)^{3/2}\big(\dd\rho^2 + \e^{6\varphi}\e^{-4\rho}(\dd x^2 + \dd y^2)\big)\,,\\
A^3 =&\, \f12 \big[(\partial_x\varphi) \dd y - (\partial_y\varphi) \dd x\big]\,,\\
\e^{4\alpha} =& \partial_\rho\varphi\,,
\end{split}
\ee
where the index $3$ on the gauge field indicates that we are turning only the $\U(1)$ Cartan generator of the $\SU(2)$ gauge group. As before, $\varphi$ satisfies  a single partial differential equation
\be\label{6Dmaster}
\triangle \varphi +\e^{6\varphi}\e^{-4\rho}\big(\partial_\rho^2 \varphi + 3 (\partial_\rho \varphi)^2 - 3 \partial_\rho \varphi \big)=0\,.
\ee
Notice that the structure of this PDE is somewhat different from the corresponding PDEs in four \eqref{4Dmaster} and five dimensions \eqref{5Dmaster}. This difference can be traced to the presence of the extra scalar field in the six-dimensional supergravity theory.

\subsubsection{The constant curvature black 2-brane}
\begin{figure}
\centering
\includegraphics[width=0.75\textwidth]{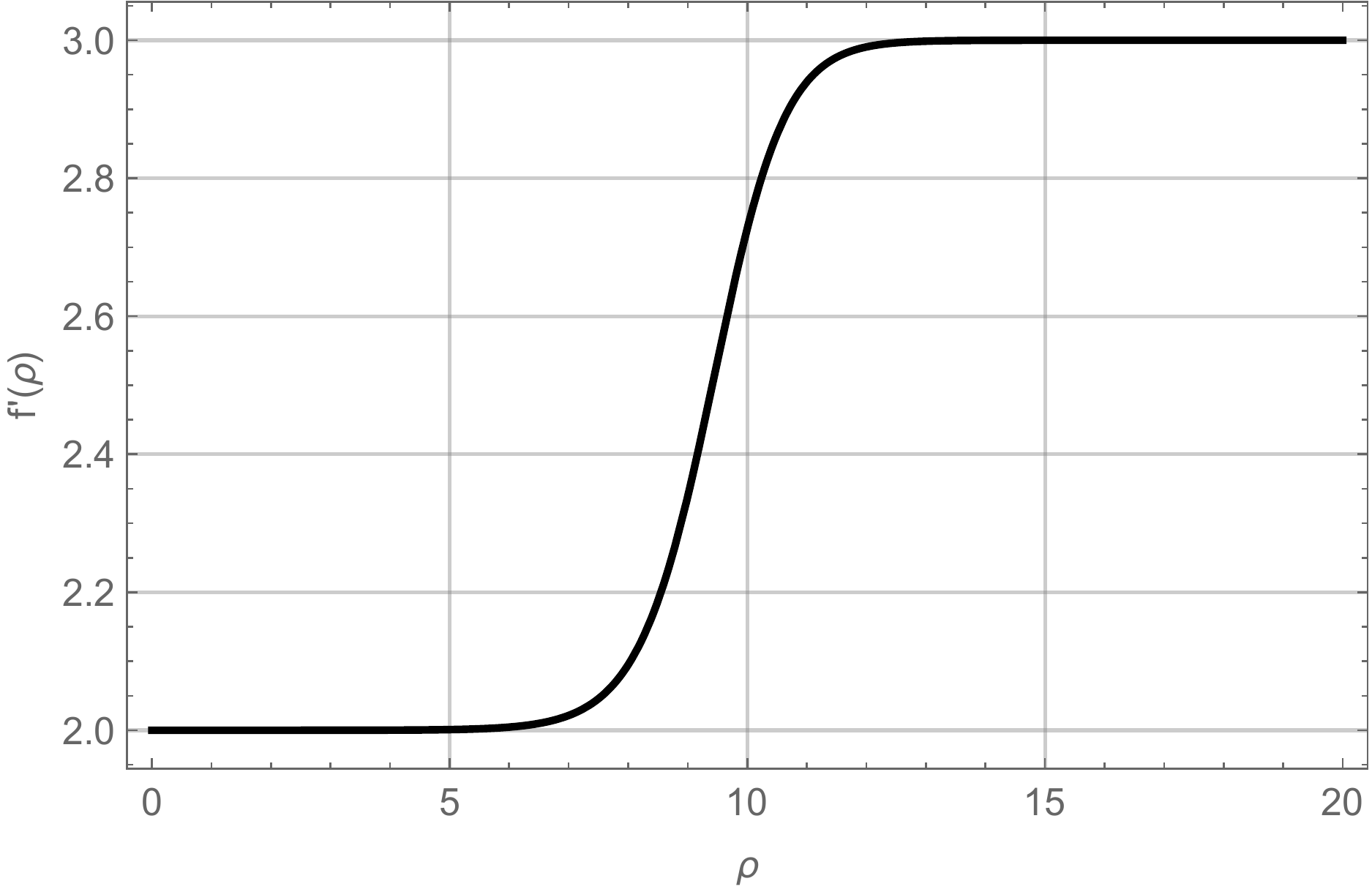}
\caption{\label{AdS6AdS4}Numerical solution of \eqref{6DLiouville} for $\mathfrak{f}(\rho)$ obtained by imposing the AdS$_4$ boundary conditions as defined by \eqref{6DFixedPointAnsatz} and \eqref{6DFixedPoints}. The figure shows $\mathfrak{f}'(\rho)$ which smoothly interpolates between the AdS$_4$ value $2$ and the AdS$_6$ value $3$.}
\end{figure}

A simple solution of \eqref{6Dmaster} corresponding to a constant curvature black brane is obtained by assuming a separable solution of the form 
\be
\varphi = \f13\big( \mathfrak{f}(\rho) + g(x,y)\big)\,,
\ee
where
\be\label{6DLiouville}
\triangle g + \kappa\,\e^{2g} =0 \,,\qquad \e^{2\mathfrak{f}-4\rho}\big(\mathfrak{f}''+\mathfrak{f}'(\mathfrak{f}'-3)\big)=\kappa\,,
\ee
and $\kappa$ is the curvature of the Riemann surface \eqref{RiemannSurfaceMetric}. The equation for $\mathfrak{f}$ in \eqref{6DLiouville} does not admit a general analytic solution. However there are two special solutions which are singled out by the fact that the scalar $\alpha$ takes a constant value. These solutions take the simple form
\be\label{6DFixedPointAnsatz}
\mathfrak{f} = c_1 \rho + c_2\,,\qquad c_1 \ne 0\,,
\ee
where $c_{1,2}$ are undetermined constants. Inserting \eqref{6DFixedPointAnsatz} into \eqref{6DLiouville} we find two possible solutions:
\be\label{6DFixedPoints}
\text{AdS$_6$:}\, c_1 = 3\,,\quad \kappa=0\,,\qquad \text{AdS$_4$:}\, c_1 = 2\,,\quad c_2 = -\f12 \log 2\,,\quad \kappa=-1\,.
\ee
The first solution, as the name indicates, is simply the maximally supersymmetric AdS$_6$ vacuum solution of the theory. The second solution corresponds to the AdS$_4$ near horizon region of a constant curvature black brane solution found in \cite{Nunez:2001}. This AdS$_4\times \Sigma_{\mathfrak{g}}$ solution exists only for $\kappa=-1$ and has the following  metric
\be
\dd s_6^2 =  \Big(\f2{27}\Big)^{3/2} \Big[2\dd s_{\text{AdS}_4}^2 + \dd s_{\Sigma_{\mathfrak{g}}}^2\Big]\,,
\ee
The full black brane solution which interpolates between the AdS$_4$ near horizon region and the asymptotically locally AdS$_6$ metric in the UV region can be constructed numerically and is displayed in Figure~\ref{AdS6AdS4}. Note that for large $\rho$ the effect of the curvature of the Riemann surface is negligible and we recover the AdS$_6$ behavior of $\mathfrak{f}(\rho)$ in \eqref{6DFixedPoints}.
%

\subsubsection{Perturbative analysis}
Since we do not have a complete analytic constant curvature black brane solution of \eqref{6DLiouville} we cannot repeat the details of the four- and five-dimensional analysis in Sections \ref{subsubsec:4Dperturbations} and \ref{subsubsec:5Dperturbations}. Nevertheless, we can still perturb the numerical solution displayed in Figure~\ref{AdS6AdS4} by employing the Ansatz in \eqref{5DperturbationsAnsatz} where the functions $\mathfrak{f}$ and $g$ satisfy \eqref{6DLiouville} and the eigenmodes $Y_{n}$ satisfy \eqref{modeExpansion}. Using this Ansatz as well as \eqref{6DLiouville} we can linearize the PDE in \eqref{6Dmaster} to find the following differential equation for the functions $\varphi_n$:
\be\label{6DperturbationsODE}
\Big[-(\mu_n+2)+\e^{2\mathfrak{f}-4\rho}\big((2\mathfrak{f}'-3)\partial_\rho + \partial_\rho^2\big)\Big]\varphi_n(\rho) =0\,.
\ee
While we cannot solve this equation without having an analytic expression for $\mathfrak{f}(\rho)$ we can still extract useful information from it by focusing on the IR region at $\rho \to -\infty$ given by the solution for $\mathfrak{f}(\rho)$ in \eqref{6DFixedPoints}. Using this solution in \eqref{6DperturbationsODE}  we find a simple ODE  with an unique regular solution 
\begin{equation}
\varphi_n = a_n\rme^{\gamma_n \rho}, \hspace{0.5cm} \gamma_n = \frac12 \big(-1 + \sqrt{17 + 8\mu_n}\big) >0\,.
\end{equation}
As in four and five dimensions we observe that the perturbations of the Riemann surface metric away from the constant curvature one vanish as we approach the IR region at  $\rho\to -\infty$. Therefore, despite the lack of an analytic solution to \eqref{6DFixedPoints} we can convincingly establish the uniformizing behavior for the metric perturbations in the IR near horizon region.

\subsubsection{UV analysis}
The UV analysis at $\rho \to \infty$ of \eqref{6Dmaster} can be performed similarly to the previous cases. The general UV behaviour of the function $\varphi$ is
\be\label{eq:UVexp6d}
\varphi =\rho + \f13 \hat{g}(x,y) + \sum_{n\ge 2} \e^{-n\rho} g_{n}(x,y)\,,
\ee
where $\hat{g}$ determines the metric on the Riemann surface and is not constrained in the UV. As in previous cases, we discover a second function, namely $v\equiv g_3$, that is unconstrained in the UV region . All other functions $g_n$ with $n\ne 3$ in \eqref{eq:UVexp6d} are related algebraically to (derivatives of) $\hat{g}(x,y)$ and $v(x,y)$ when equation \eqref{6Dmaster} is solved order by order in $\e^{-\rho} \to 0$. We can then proceed to employ similar arguments to the one in Section~\ref{subsubsec:4dUV} to conclude that the PDE in \eqref{6Dmaster} leads to a uniformization flow for the metric on the Riemann surface which approaches the constant curvature metric near the regular AdS$_4$ near horizon region.

 \subsection{Seven dimensions}

For completeness we present here also the universal black brane in the minimal seven-dimensional gauged supergravity of \cite{Townsend:1983kk}. This case was treated in Section 3.2 of \cite{Anderson:2011cz} to which we refer for a detailed derivation of the BPS configuration.\footnote{The analysis in \cite{Anderson:2011cz} was performed in the maximal seven-dimensional $\SO(5)$ gauged supergravity of which the minimal theory in \cite{Townsend:1983kk} is a consistent truncation.} The bosonic field content is the same as in six dimensions, namely the metric, an $\SU(2)$ gauge field and a real scalar.

The static BPS black brane solutions are given by\footnote{When comparing with \cite{Anderson:2011cz}, note that we have set $m=2$ and $\varphi_\text{there} = 8\varphi_\text{here}-6\rho + 2\log(y)$.}
\be
\begin{split}
\dd s_7^2  =&\, \f{\e^{2\rho}\,\dd s_{\mathbf{R}^{1,3}}^2}{(\partial_\rho\varphi)^{2/5}} + (\partial_\rho\varphi)^{8/5}\Big(\dd\rho^2 + \e^{8\varphi}\e^{-6\rho}(\dd x^2 + \dd y^2)\Big)\,,\\
A^3 =&\, \f12 \big[(\partial_x\varphi) \dd y - (\partial_y\varphi) \dd x\big]\,,\\
\e^{-5\phi} =& \partial_\rho\varphi\,,
\end{split}
\ee
where $\varphi$ satisfies 
\be\label{7Dmaster}
\triangle \varphi +\e^{8\varphi}\e^{-6\rho}\Big(\partial_\rho^2 \varphi + 4 (\partial_\rho \varphi)^2 - 4 \partial_\rho \varphi \Big)=0\,.
\ee
Note the similarity between this equation and the six-dimensional one in \eqref{6Dmaster}.

\subsubsection{The constant curvature black 3-brane}
\begin{figure}
\centering
\includegraphics[width=0.75\textwidth]{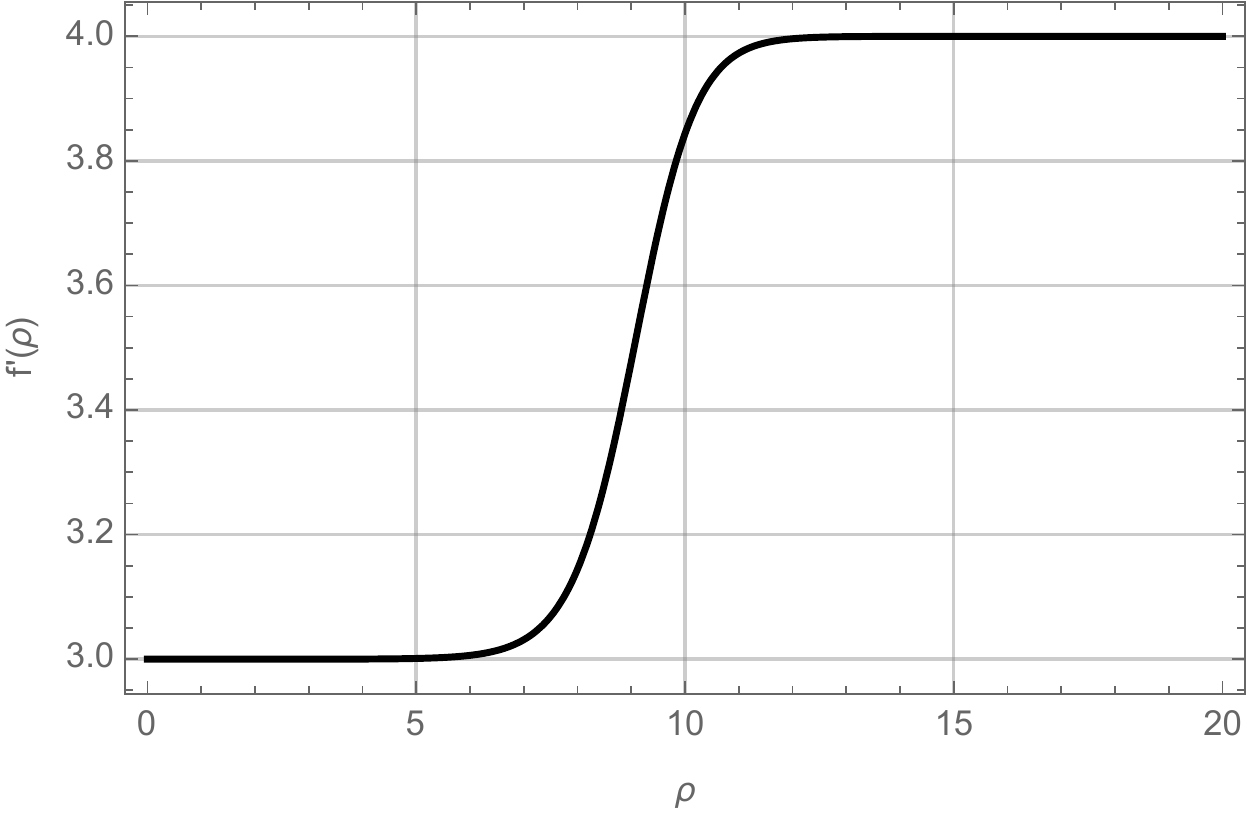}
\caption{\label{AdS7AdS5}Numerical solution of \eqref{7DLiouville} for $\mathfrak{f}(\rho)$ obtained by imposing the AdS$_5$ boundary conditions as defined by \eqref{7DFixedPointAnsatz} and \eqref{7DFixedPoints}. The figure shows $\mathfrak{f}'(\rho)$ which smoothly interpolates between the AdS$_5$ value $3$ and the AdS$_7$ value $4$.}
\end{figure}

As is familiar by now we look for a simple solution of \eqref{7Dmaster} by assuming a separable form of the function $\varphi$ 
\be
\varphi = \f14\big( \mathfrak{f}(\rho) + g(x,y)\big)\,,
\ee
where
\be\label{7DLiouville}
\triangle g + \kappa\,\e^{2g} =0 \,,\qquad \e^{2\mathfrak{f}-6\rho}\big(\mathfrak{f}''+\mathfrak{f}'(\mathfrak{f}'-4)\big)=\kappa\,,
\ee
and $\kappa$ is the curvature of the Riemann surface \eqref{RiemannSurfaceMetric}. The equation for $\mathfrak{f}$ in \eqref{7DLiouville} does not admit a general analytic solutions, however we can again find simple solutions with a constant scalar field which take the form
\be\label{7DFixedPointAnsatz}
\mathfrak{f} = c_1 \rho + c_2\,,\qquad c_1 \ne 0\,,
\ee
where $c_{1,2}$ are undetermined constants. Using \eqref{7DFixedPointAnsatz} in \eqref{7DLiouville} we obtain two solutions:
\be\label{7DFixedPoints}
\text{AdS$_7$:}\, c_1 = 4\,,\quad \kappa=0\,,\qquad \text{AdS$_5$: }\, c_1 = 3\,,\quad c_2 = -\f12 \log 3\,,\quad \kappa=-1\,.
\ee
The first solution is the AdS$_7$ supersymmetric vacuum solution of the gauged supergravity. The AdS$_5\times \Sigma_{\mathfrak{g}}$ solution with $\kappa=-1$ represents the near horizon geometry of the constant curvature black brane studied in \cite{Maldacena:2000mw}. In this case the seven-dimensional metric is given by
\be
\dd s_7^2 =  \f18\Big(\f{27}{2}\Big)^{1/5} \Big[3\dd s_{\text{AdS}_5}^2 + \dd s_{\Sigma_{\mathfrak{g}}}^2\Big]\,.
\ee
A numerical solution interpolating between the AdS$_5$ and the AdS$_7$ region of the black brane geometry is displayed in Figure \ref{AdS7AdS5}. Note that for large $\rho$ the effect of the curvature of the Riemann surface is negligible and we recover the AdS$_7$ behaviour of $\mathfrak{f}(\rho)$ in \eqref{7DFixedPoints}. 

\subsubsection{Perturbative analysis}
In the absence of a complete analytic solution of \eqref{7DLiouville} we cannot find explicitly the perturbations around the constant curvature black brane black. However, it is still possible to perturb the solution displayed in Figure~\ref{AdS7AdS5} using the Ansatz in \eqref{5DperturbationsAnsatz} where the functions $\mathfrak{f}$ and $g$ satisfy \eqref{7DLiouville} and $Y_{n}$ are defined by \eqref{modeExpansion}. We can then linearize the PDE in \eqref{7Dmaster} and find the following differential equation for the perturbations
\be\label{7DperturbationsODE}
\Big[-(\mu_n+2)+\e^{2\mathfrak{f}-6\rho}\big((2\mathfrak{f}'-4)\partial_\rho + \partial_\rho^2\big)\Big]\varphi_n(\rho) =0\,.
\ee
While we cannot solve this equation without an analytic expression for $\mathfrak{f}(\rho)$ we can extract the behavior of the perturbations in the IR region. to do this we use the IR solution for $\mathfrak{f}(\rho)$ given in \eqref{7DFixedPoints} which reduces \eqref{7DperturbationsODE} to a simple ODE  with a unique regular solution
\begin{equation}\label{eq:7dmingamansol}
\varphi_n = a_n\rme^{\gamma_n \rho}, \hspace{0.5cm} \gamma_n = -1 + \sqrt{7 + 3\mu_n}>0\,.
\end{equation}
As in all previous cases we observe that the perturbations of the constant curvature metric on the Riemann surface vanish as we approach the AdS$_5$ near horizon region at $\rho\to -\infty$. This is a manifestation of the uniformization behavior exhibited in \cite{Anderson:2011cz}.

\subsubsection{UV analysis}
The UV analysis at $\rho \to \infty$ of the equation \eqref{7Dmaster} can be performed similarly to previous cases and was discussed in detail in \cite{Anderson:2011cz}. The UV behavior of the function $\varphi$ is
\be
\varphi =\rho + \f14 \hat{g}(x,y) + \sum_{n\ge 2}\sum_{m=0}^{n-1} \e^{-2n\rho}\rho^{m} g_{n,m}(x,y)\,,
\ee
where $\hat{g}(x,y)$ determines the metric on the Riemann surface and is not constrained in the UV. As in previous cases, we find another function, $v\equiv g_{2,0}$, that is unconstrained in the UV and which has to be chosen appropriately in terms of $\hat{g}(x,y)$ to ensure regular solutions in the IR region. All other functions $g_{n,m}$ are related algebraically to (derivatives of) $\hat{g}(x,y)$ and $v(x,y)$ when equation \eqref{7Dmaster} is solved order by order in $\e^{-\rho} \to 0$. In addition to this perturbative evidence for the uniformization behavior of the metric on the Riemann surface a global existence proof of these uniformizing solutions of \eqref{7Dmaster} was provided in \cite{Anderson:2011cz}.

\section{Black holes with scalar hair}
\label{sec:STU}

After we have established the holographic uniformization principle for the black brane solutions of gauged supergravity theories consisting only of the gravity multiplet it is natural to generalize this analysis by coupling these theories to matter multiplets. We focus our attention on matter coupled gauged supergravity models in four, five, and seven dimensions, which arise as consistent truncation of ten- or eleven-dimensional supergravity. These so-called STU models share several common features and which are reflected in similarities in the analysis below. In particular, as discussed in \cite{Cvetic:1999xp}, these theories arise as a Kaluza-Klein reduction on a sphere from ten or eleven dimensions where one restricts to the gauge field, and accompanying dilatonic scalar fields, associated with the Cartan subalgebra of the isometries of the sphere. Using the results in \cite{Cvetic:1999xp} we can therefore uplift all solutions discussed below to backgrounds in type IIB or eleven-dimensional supergravity.

The BPS equations for these STU models are more complicated than the ones in the minimal supergravity discussed in the previous section. In particular we find that the BPS equations reduce to a coupled system of PDEs for functions of three variables. We study these equations both in the IR and the UV region and establish that despite this more involved structure the solutions still exhibit the expected uniformization behavior. The solutions we study should be viewed as generalizations of the constant curvature black branes in seven-dimensions \cite{Bah:2011vv,Bah:2012dg}, black strings in five-dimensions \cite{Benini:2012cz,Benini:2013cda}, and black holes in four dimensions \cite{Gauntlett:2001qs,Cacciatori:2009iz,Benini:2015eyy}.

 \subsection{Seven-dimensional STU model}

The solutions of interest are constructed in the $\U(1)^2$ invariant sector of the maximal $\SO(5)$ gauged supergravity of \cite{Pernici:1984xx}. This truncation was studied in \cite{Liu:1999ai} and was used to construct supersymmetric black brane solutions arising from M5-branes wrapped on $\Sigma_{\mathfrak{g}}$ in \cite{Bah:2012dg,Bah:2011vv}. Following our general strategy we will generalize the solutions of \cite{Bah:2012dg,Bah:2011vv} by allowing for the metric on $\Sigma_{\mathfrak{g}}$ to be arbitrary. 

The bosonic sector of the supergravity truncation we study consists of the metric, two $\U(1)$ gauge fields $A^1_\mu$ and $A^2_\mu$ and two real scalar fields, $\alpha$ and $\beta$. The Lagrangian of this model is, see \cite{Maldacena:2000mw,Liu:1999ai},
\begin{equation}\begin{split}
\mathcal{L} = &R - 5\partial (\alpha + \beta)^2 -\partial(\alpha - \beta)^2 - \rme^{-4\alpha}(F^1_{\mu\nu})^2 -\rme^{-4\beta}(F^2_{\mu\nu})^2\\ &- 2(-8\rme^{2(\alpha + \beta)}-4\rme^{-2\alpha - 4\beta}-4\rme^{-4\alpha-2\beta}+\rme^{-8(\alpha+\beta)})\,,    
\end{split}
\end{equation}
where we are using conventions in which the AdS$_7$ length scale is set to 1. The supersymmetry variations of this model take the form
\begin{equation}\label{eq:7dsusyvar}
\begin{split}
\delta \psi_\mu = \Big[ &\nabla_\mu + 2(A^1_\mu \Gamma^{12} + A^2_\mu \Gamma^{34}) + \frac12 \rme^{-4(\alpha + \beta)}\gamma_\mu + \frac12 \gamma_\mu \gamma^\nu \partial_\nu (\alpha + \beta)\\ &+ \frac12 \gamma^\nu (\rme^{-2\alpha}F^1_{\mu\nu}\Gamma^{12}+ \rme^{-2\beta}F^2_{\mu\nu}\Gamma^{34})\Big] \epsilon\,,\\
\delta \chi^{(1)} = &\Big[ \frac12 (\rme^{2\alpha}-\rme^{-4(\alpha+\beta)})-\frac14 \gamma^\mu \partial_\mu(3 \alpha+2\beta) -\frac18 \rme^{-2\alpha} \gamma^{\mu\nu}F^1_{\mu\nu}\Gamma^{12}\Big]\epsilon\,,\\
\delta \chi^{(2)} = &\Big[ \frac12 (\rme^{2\beta}-\rme^{-4(\alpha+\beta)})-\frac14 \gamma^\mu \partial_\mu (2\alpha+3\beta) -\frac18 \rme^{-2\beta} \gamma^{\mu\nu}F^2_{\mu\nu}\Gamma^{34} \Big]\epsilon\,.
\end{split}
\end{equation}

We are interested in static supersymmetric solutions of this model captured by the following Ansatz
\begin{equation}\label{eq:7dstuAnsatz}
\begin{split}
\rmd{s}^2 &= \rme^{2f(r,x,y)}\dd s_{\mathbf{R}^{1,3}}^2 +\rme^{2h(r,x,y)}\rmd{r}^2 +\rme^{2g(r,x,y)}(\rmd{x}^2 + \rmd{y}^2)\,,\\
A^a &= A^a_x(r,x,y) \rmd{x} + A^a_y(r,x,y) \rmd{y}\,,\\
\alpha &= \alpha(r,x,y), \qquad  \beta = \beta(r,x,y)\,.
\end{split} 
\end{equation}
As discussed in Appendix~\ref{4Dapp} it can be shown that the coordinate dependence of the supersymmetry parameter in \eqref{eq:7dsusyvar} takes the simple form $\epsilon = \rme^{f/2}\epsilon_0$ where $\epsilon_{0}$ is a constant spinor. In addition the spinor obeys the following projectors
\begin{equation}
\gamma_{\Hat{r}}\epsilon_0 = \epsilon_0, \hspace{0.5cm}\gamma_{\Hat{x}\Hat{y}}\epsilon_0 = i \epsilon_0, \hspace{0.5cm} \Gamma^{12}\epsilon_0 = \Gamma^{34}\epsilon_0 = i\epsilon_0\,.
\end{equation}
Here $\Gamma$ are $\SO(5)$ gamma matrices and we have denoted tangent space indices with a hat.

With this Ansatz at hand the spin-1/2 variations in \eqref{eq:7dsusyvar} variations reduce to the following PDEs
\begin{eqnarray}
\label{7dgen: dil1}\rme^{-h}\partial_r (3\alpha + 2\beta) -2(\rme^{2\alpha}-\rme^{-4(\alpha+\beta)})-\rme^{-2\alpha-2g}F^1_{xy}=0\,,\\
\label{7dgen: dil2}(\partial_x + i\partial_y)(3\alpha + 2\beta)-i\rme^{-2\alpha-h}(F^1_{rx}+ i F^1_{ry})=0\,,\\
\label{7dgen: dil3}\rme^{-h}\partial_r (2\alpha + 3\beta) -2(\rme^{2\beta}-\rme^{-4(\alpha+\beta)})-\rme^{-2\beta-2g}F^2_{xy}=0\,,\\
\label{7dgen: dil4}(\partial_x + i\partial_y)(2\alpha + 3\beta)-i\rme^{-2\beta-h}(F^2_{rx}+ i F^2_{ry})=0\,.
\end{eqnarray}
The $t$-component of the spin-3/2 supersymmetry variation lead to two differential constraints which determine that two combinations of the metric functions and the scalar fields in the model are independent of the $x,y$ coordinates on $\Sigma_{\mathfrak{g}}$ 
\begin{equation}\label{7dgen: rho}
    (\partial_x + i\partial_y)(f+\alpha + \beta)=0, \hspace{0.5cm} \partial_r (f+\alpha+\beta) + \rme^{h-4(\alpha+\beta)}=0\,.
\end{equation}
The other components of the spin-3/2 variation reduce to the differential equation
\begin{equation}\label{eq:spin327dr}
\partial_r(g-4\alpha-4\beta)-3\rme^{h-4(\alpha+\beta)}+2\rme^{h+2\alpha}+2\rme^{h+2\beta}=0\,,
\end{equation}
and the following expression for the gauge fields
\begin{equation}\label{eq:7dspin32gauge}
A^1_x + A^2_x = -\frac14\partial_y(g-4\alpha - 4\beta)\,, \qquad A^1_y + A^2_y = \frac14\partial_x(g-4\alpha - 4\beta)\,.
\end{equation}
The equations in \eqref{7dgen: dil1}-\eqref{eq:7dspin32gauge} are all constraints imposed on the bosonic fields of the Ansatz in \eqref{eq:7dstuAnsatz}.

To analyze these equations it is helpful to use \eqref{7dgen: rho} to define a new radial coordinate $\rho$ via $\rmd{\rho} = - \rme^{h-4(\alpha + \beta)} \rmd{r}$. This in turn implies that $f+\alpha + \beta = \rho$. It proves useful also to define the following combinations 
\begin{equation}
\varphi \equiv 2g - 8\alpha - 8\beta\,, \qquad \xi \equiv 6\alpha + 4\beta\,, \qquad \chi \equiv 4\alpha + 6\beta\,.
\end{equation}
Equipped with this we arrive at the following set of BPS equations which ensure that the configuration in \eqref{eq:7dstuAnsatz} preserves 1/4 of the maximal supersymmetry.
\begin{eqnarray}
\label{7dgen: 1}\partial_\rho \varphi = 4\Big(\rme^\xi + \rme^\chi-\frac32\Big)\,,\\
\label{7dgen: 6}\triangle\varphi =8(F^1_{xy}+ F^2_{xy})\,,\\
\label{7dgen: 2}\partial_\rho \xi + 4(\rme^\xi -1) + 2 \rme^{-\varphi-\xi}F^1_{xy}=0\,,\\
\label{7dgen: 3}\partial_\rho \chi + 4(\rme^\chi -1) + 2 \rme^{-\varphi-\chi}F^2_{xy}=0\,,\\
\label{7dgen: 41}(\partial_x + i\partial_y)\rme^\xi = 2i \rme^{-h+4\alpha+4\beta}(F^1_{rx}+iF^1_{ry})\,,\\
\label{7dgen: 5}(\partial_x + i\partial_y)\rme^\chi = 2i \rme^{-h+4\alpha+4\beta}(F^2_{rx}+iF^2_{ry})\,.
\end{eqnarray}
This system of equations can be further reduced to only two coupled PDEs given by
\begin{equation}\label{7dgen: fin1}
\partial_\rho\lbrack\rme^{\varphi + \chi}(\partial_\rho \chi + 4(\rme^\chi-1)) \rbrack + \triangle\rme^\chi=0\,,
\end{equation}
and
\begin{equation}\label{7dgen: fin2}
\rme^{\varphi} \partial^2_\rho \varphi + \rme^{\varphi}(\partial_\rho\varphi)^2+ 32 \rme^{\varphi + 2\chi}-48\rme^{\varphi + \chi}+ 12\rme^{\varphi}-8\rme^{\varphi}\partial_\rho\varphi (\rme^\chi-1)+ \triangle\varphi=0\,.
\end{equation}
To derive these equations we made use of the Bianchi identity to eliminate the field strengths from \eqref{7dgen: 3} and \eqref{7dgen: 5} and reorganized appropriately the equations in \eqref{7dgen: 1}-\eqref{7dgen: 3}.

Given a solution for $\chi$ and $\varphi$ of the two coupled PDEs \eqref{7dgen: fin1} and \eqref{7dgen: fin2} one can obtain all fields in the background \eqref{eq:7dstuAnsatz} using only derivatives and algebraic relations. Therefore to find supersymmetric backgrounds described by \eqref{eq:7dstuAnsatz} it is sufficient to focus on solutions of \eqref{7dgen: fin1} and \eqref{7dgen: fin2}. Notice that this is considerably more complicated than the situation in minimal seven-dimensional supergravity where similar solutions were described by the single PDE in \eqref{7Dmaster}.

\subsubsection{IR analysis}

We start with the analysis of the BPS equations in the IR region where we look for regular AdS$_5$ solutions. We can solve \eqref{7dgen: 1}-\eqref{7dgen: 6}, by making the Ansatz
\begin{equation}
\xi_{\rm IR} = \log\xi_0, \qquad \chi_{\rm IR} = \log\chi_0, \hspace{0.5cm}\varphi_{\rm IR} = \log\varphi_0-2 g(x,y)\,.
\end{equation}
We find then that $g(x,y)$ obeys the Liouville equation in \eqref{7DLiouville} and thus describes a Riemann surface of constant curvature $\kappa$. We focus on the case $\kappa \neq 0$ and use the notation of \cite{Bah:2012dg} for the field strengths of the gauge fields\footnote{The case $\kappa=0$ can be analyzed analogously following Appendix C of \cite{Bah:2012dg}.}
\begin{equation}
F^1_{xy}= p\rme^{-2g(x,y)}, \hspace{0.5cm} F^2_{xy}= q\rme^{-2g(x,y)}\,.
\end{equation}
Supersymmetry imposes a relation between the two magnetic fluxes which can be conveniently expressed by rewriting them in terms of a single parameter $z$ 
\begin{equation}
p = -\frac{\kappa(1+z)}{8}\,, \qquad   q = -\frac{\kappa(1-z)}{8}\,.
\end{equation}
Note that $z$ should be quantized such that $z(\mathfrak{g}-1) \in \mathbb{Z}$.

The two equations \eqref{7dgen: 1}-\eqref{7dgen: 6} then reduce to
\begin{eqnarray}
4(\xi_0-1) + \frac2{\xi_0\varphi_0}p=0\,,\\
4(\chi_0-1) + \frac2{\chi_0\varphi_0}q=0\,,\\
\xi_0 + \chi_0 -\frac32 = 0\,.
\end{eqnarray}
We thus find the IR solutions
\begin{equation}
\xi_0 = \frac{1+3z+\kappa\sqrt{1+3z^2}}{4z}\,,\quad\chi_0 = \frac{3z-1-\kappa\sqrt{1+3z^2}}{4z}, \quad\varphi_0 = \frac{\sqrt{1+3z^2}-\kappa}{6}\,.\notag
\end{equation}
We have thus arrived at the family of AdS$_5$ solutions with a constant curvature metric on $\Sigma_{\mathfrak{g}}$ found in \cite{Bah:2012dg}. Note that to ensure regularity of the solution we need to restrict  $|z|>1$ for $\kappa = 1$. There is no upper or lower bound on the value of $z$ for $\kappa=-1$.
 
Our goal now is to expand \eqref{7dgen: fin1} and \eqref{7dgen: fin2} around these IR AdS$_5$ solutions in the same way as before, by writing $\varphi = \varphi_{\rm IR}+ \tilde{\varphi}$ and expanding the small perturbation $\tilde{\varphi}$ in eigenfunctions of the constant curvature Laplacian as follows
\begin{equation}
    \varphi = \varphi_{\rm IR} +  \sum^{\infty}_{n=0}\tilde{\varphi}_n (\rho) Y_n(x,y), \qquad \triangle_g Y_n = - \mu_n Y_n.
\end{equation}
We perform a similar expansion for the function $\chi$ and use \eqref{7dgen: fin1} and \eqref{7dgen: fin2} to obtain the following system of linear equations 
\begin{eqnarray}
\partial^2_\rho \tilde{\chi}_n + 4(2\chi_0-1)\partial_\rho\tilde{\chi}_n+4(\chi_0-1)\partial_\rho\tilde{\varphi}_n-\mu_n \varphi^{-1}_0\tilde{\chi}_n=0\,,\\
\partial^2_\rho \tilde{\varphi}_n - 8(\chi_0 -1)\partial_\rho \tilde{\varphi}_n + (64\chi^2_0-48\chi_0)\tilde{\chi}_n-(\mu_n - 2\kappa)\varphi^{-1}_0\tilde{\varphi}_n=0\,.
\end{eqnarray}
A regular solution of these takes the form
\begin{equation}\label{eq:phinxchinsol7d}
\begin{pmatrix}\tilde{\varphi}_n\\ \tilde{\chi}_n \end{pmatrix} = \rme^{\gamma_n \rho}\begin{pmatrix}A\\B \end{pmatrix}\,.
\end{equation}
where the real constants $A$ and $B$ are determined by solving the following matrix equation
\begin{equation}\label{eq:gammaMat7d}
\begin{pmatrix}4(\chi_0-1)\gamma_n & \gamma^2_n +4(2\chi_0-1) -\mu_n \varphi^{-1}_0 \\ \gamma^2_n - 8(\chi_0 -1)\gamma_n -(\mu_n - 2\kappa)\varphi^{-1}_0 & 64\chi^2_0-48\chi_0 \end{pmatrix} \begin{pmatrix}A\\B \end{pmatrix} = 0\,.
\end{equation}
The vanishing of the determinant of the matrix \eqref{eq:gammaMat7d} leads to an equation for the constants $\gamma_n$. Importantly there are always two positive roots of this equation  for $\kappa=-1,1$ given explicitly by
\begin{equation}
\gamma_n = -1 + \sqrt{\frac{-\kappa+7\sqrt{1+3z^2}+6\mu_n \pm 6 \sqrt{1+3z^2 + 2(\kappa+\sqrt{1+3z^2})\mu_n}}{-\kappa+\sqrt{1+3z^2}}}\,.
\end{equation}
We therefore conclude that indeed for each choice of eigenmode on $\Sigma_{\mathfrak{g}}$ there is a regular solution of the form \eqref{eq:phinxchinsol7d} which describes a small deviation from the constant curvature solution. Note that for $z=0$ and $\kappa=-1$, the largest solution for $\gamma_n$ reduces to the one in  \eqref{eq:7dmingamansol}.

\subsubsection{UV analysis}
In the UV region at $\rho \to \infty$ the BPS equations \eqref{7dgen: fin1} and \eqref{7dgen: fin2} can be solved systematically order by order for any metric on the Riemann surface. The solution is given by
\begin{equation}\begin{split}
    \varphi &= 2 \rho + \varphi_0 + \varphi_2 \rme^{-2\rho} + \varphi_4 \rme^{-4\rho} + \varphi_{4,1}\rme^{-4\rho} \rho + \text{higher order,}\\
    \chi &= \chi_2 \rme^{-2\rho} + \chi_4 \rme^{-4\rho} + \chi_{4,1}\rme^{-4\rho} \rho + \text{higher order,}\end{split}
\end{equation}
where the functions $\varphi_0(x,y)$, $\varphi_4(x,y)$, $\chi_2(x,y)$, and $\chi_4(x,y)$ can be chosen freely. The rest of the UV expansion functions are determined in terms of these four. For example, we find
\begin{equation}\begin{split}
        \varphi_2 &= \frac14 \rme^{-\varphi_0}\triangle\varphi_0\,,\\
       \varphi_{4,1} &=-8\chi^2_2 -4\chi_2\varphi_2 - \frac14 \rme^{-\varphi_0}\triangle\varphi_2\,,\\
       \chi_{4,1} &= 4\chi^2_2 +2\chi_2\varphi_2 -\frac12 \rme^{-\varphi_0}\triangle\chi_2\,.
\end{split}
\end{equation}
Taken together, the UV and the IR expansion discussed above strongly suggest that, for arbitrary values of $z$ and $\kappa$, there indeed exist full nonlinear solutions of the BPS equations of this seven-dimensional model which interpolate between the IR and UV region and uniformize the arbitrary UV metric on the Riemann surface. For $z=0,1$ and $\kappa=-1$ this statement was proven rigorously in \cite{Anderson:2011cz}.

 \subsection{Five-dimensional STU model}

Here we study the well-known five-dimensional STU model of $\mathcal{N}=2$ five-dimensional gauged supergravity which arises from a consistent truncation of type IIB supergravity on $S^5$, see \cite{Cvetic:1999xp}. The bosonic fields of this theory are the metric, three Abelian gauge fields and two real scalars. The Lagrangian for these fields is given by, see for instance \cite{Maldacena:2000mw,Benini:2013cda},
\begin{equation}
\begin{split}
    \mathcal{L} = &R - \frac12 (\partial_\mu \phi_1)^2 - \frac12 (\partial_\mu \phi_2)^2 + 12 \sum^3_{a=1} X_a\\ &- \frac94 \sum^3_{a=1} X_a^2 (F^a_{\mu\nu})^2 + \frac14 \epsilon^{\mu\nu\alpha\beta\sigma}F^1_{\mu\nu}F^2_{\alpha\beta}A^3_\sigma\,.
    \end{split}
\end{equation}
The three gauge fields, $A^a$, correspond to the Cartan generators of the $\SO(6)$ isometry group of $S^5$. It is also useful to define the ``sections'' $X_a$ which are related to the scalars $\phi_1$ and $\phi_2$ via 
\begin{equation}
X^1 = \rme^{-\frac{\phi_1}{\sqrt{6}}-\frac{\phi_2}{\sqrt{2}}}, \hspace{0.5cm} X^2 = \rme^{-\frac{\phi_1}{\sqrt{6}}+\frac{\phi_2}{\sqrt{2}}}, \hspace{0.5cm} X^3 = \rme^{2\frac{\phi_1}{\sqrt{6}}}\,,
\end{equation}
as well as their inverse
\begin{equation}
X_a = \frac{1}{3 X^a}\,.
\end{equation}
The fermionic supersymmetry variations of the supergravity are those of the two dilatini $\chi_{(j)}$ and the gravitino $\psi$ and read
\begin{equation}\label{eq:5dsusyvarSTU}
\begin{split}
    \delta \psi_\mu&= \Big[ \nabla_\mu + \frac{i}{8}X_a(\gamma^{\nu \rho}_\mu-4\delta^\nu_\mu \gamma^\rho)F^a_{\nu\rho} +\frac12 X^a V_a \gamma_\mu - \frac{3i}{2}V_a A^a_{\mu}\Big] \epsilon\,, \\
    \delta \chi_{(j)} &= \Big[ \frac38 (\partial_{\phi_j}X_a) F^a_{\mu \nu}\gamma^{\mu \nu} + \frac{3i}{2}V_a \partial_{\phi_j}X^a - \frac{i}{4}\delta_{jk} \partial_\mu \phi_k \gamma^\mu \Big] \epsilon\,, 
\end{split}    
\end{equation}
where $j=1,2$ and $V_a = \frac13$. We now proceed to analyze these equations to find static supersymmetric black string solutions.

\subsubsection{BPS constraints}
The Ansatz we consider is similar to the one used to construct the supersymmetric black string solutions in \cite{Benini:2013cda,Benini:2012cz}, however we now allow for a general metric on the Riemann surface
\begin{equation}\label{eq:5dstuansatz}
\begin{split}
\rmd{s}^2 &= \rme^{2f(r,x,y)}\dd s_{\mathbf{R}^{1,1}}^2 + \rme^{2h(r,x,y)}\rmd{r}^2+\rme^{2g(r,x,y)}(\rmd{x}^2+\rmd{y}^2)\,,\\
A^a &= A^a_x(r,x,y) \rmd x + A^a_y (r,x,y) \rmd y\,, \\
\frac{\phi_1}{\sqrt{6}} &= \alpha(r,x,y), \hspace{0.5cm}\frac{ \phi_2}{\sqrt{2}}= \beta(r,x,y) \,,
\end{split}\end{equation}
Note that for convenience we have rescaled the scalar fields $\phi_{1,2}$ by suitable constants. As is familiar by now, the BPS equations of the model are then obtained after taking $\epsilon = \rme^{f/2}\epsilon_0$ and imposing the projectors
\begin{equation}\label{5dgen: proj}
\gamma_{\Hat{r}}\epsilon_0 = \epsilon_0\,, \quad \gamma_{\Hat{x}\Hat{y}}\epsilon_0 = -i \epsilon_0\,.
\end{equation}

We then find the following set of equations which ensure that the spin-1/2 supersymmetry variations in \eqref{eq:5dsusyvarSTU} are obeyed 
\be\label{5dgen: dil1}\begin{split}
 6 \partial_r \alpha + 2\rme^h (X^1 + X^2- 2X^3) + 3\rme^{h-2g}(X_a F^a_{xy}   -3 X_3 F^3_{xy}) = 0\,,\\
2 (\partial_x + i \partial_y)\alpha +  i \rme^{-h}(X_a(F^a_{rx}+ i F^a_{ry}) - 3 X_3 (F^3_{rx}+ i F^3_{ry})) = 0\,,\\
2 \partial_r \beta + 2(X^1- X^2) + 3\rme^{h-2g} ( X_1 F^1_{xy}- X_2 F^2_{xy})=0\,,\\
2 (\partial_x + i \partial_y)\beta + 3i \rme^{-h}( X_1 (F^1_{rx}+ i F^1_{ry})- X_2 (F^2_{rx}+ i F^2_{ry}))=0\,.
\end{split}\ee
The $t$- and $z$-component of the gravitino variations yield 
\be\label{5dgen: grav1}\begin{split}
6 \partial_r f + 2\rme^h\sum^3_{a=1}X^a + 3\rme^{h-2g}X_a F^a_{xy} =0\,,\\
2 (\partial_x + i \partial_y)f +  i \rme^{-h}X_a(F^a_{rx}+ i  F^a_{ry}) = 0\,,
\end{split}\ee
The other components of the gravitino variation give the following constraints on the metric functions
\begin{equation}\label{5dgen: con1}
(\partial_x + i \partial_y)(2f+h)=0\,,
\end{equation}
scalars
\begin{equation}\label{5dgen: con2}
     \partial_r (2f+g) +  \rme^{h}\sum^3_{a=1}X^a=0\,,
\end{equation}
and gauge fields
\begin{equation}\label{5dgen: gauge}
    \sum^3_{a=1}A^a_x= - \partial_y(2f+g),\qquad \sum^3_{a=1}A^a_y = \partial_x (2f+g)\,.
\end{equation}

These BPS constraints can be written more compactly by using the relation in \eqref{5dgen: con1} to define a new radial variable via $\rmd{\rho} = -\rme^{2f+h}\rmd{r}$ and then define the combinations
\begin{equation}\label{eq:5dvarphidefstu}
    \varphi \equiv 4f + 2g, \hspace{0.5cm} \xi_1 \equiv 2f + \alpha +\beta, \hspace{0.5cm} \xi_2 \equiv 2f+\alpha-\beta, \hspace{0.5cm} \xi_3 \equiv 2f-2\alpha\,.
\end{equation}
Indeed, we can now reduce the BPS constraints to the following set of equations
\begin{equation}\label{5dgen: 1}
    \partial_\rho \varphi = 2(\rme^{-\xi_1}+ \rme^{-\xi_2} + \rme^{-\xi_3})
\end{equation}
\begin{equation}\label{5dgen: 2}
    2\sum^3_{a=1}F^a_{xy}- \triangle\varphi = 0
\end{equation}
\begin{equation}\label{5dgen: 3}
    \partial_\rho \rme^{-\xi_a} + 2\rme^{-2\xi_a}+  \rme^{-\varphi}F^a_{xy}=0, \hspace{0.5cm} a =1,2,3
\end{equation}
\begin{equation}\label{5dgen: 4}
    (\partial_x + i \partial_y)\rme^{-\xi_a} = i(F^a_{rx}+iF^a_{ry})\rme^{-h-2f}, \hspace{0.5cm} a =1,2,3.
\end{equation}
A few comments are in order. We note that the three gauge fields can develop independent profiles along the radial flow. When we set all three of the gauge fields to be equal to each other we see that the two scalars fields can be consistently set to zero and we then recover the BPS equations of the minimal gauged supergravity discussed in Section~\ref{subsec:5dmin}. In the two special cases when $F^{1}=F^2 \neq0$ and $F^3=0$ or $F^1=F^2=0$ and $F^3\neq0$ we recover the 1/4-BPS and 1/2-BPS solutions studied in \cite{Anderson:2011cz}. For more general profiles of $F^{a}$ the solutions to the BPS equations above preserve 1/8 of the maximal supersymmetry.

To make progress in simplifying further the BPS equations we can invoke the Bianchi identity for \eqref{5dgen: 3} and \eqref{5dgen: 4} to obtain
\begin{equation}\label{5dgen: bi}
    \partial_\rho(\rme^\varphi (\partial_\rho \rme^{-\xi_a} + 2\rme^{-2\xi_a})) + \triangle\rme^{-\xi_a} =0, \hspace{0.5cm}\text{for }a=1,2,3
\end{equation}
The sum of the three equations \eqref{5dgen: 3} yields, in combination with \eqref{5dgen: 1}, the following equation for $\varphi$
\begin{equation}\label{5dgen: extra}
    \partial^2_\rho\varphi + 4(\rme^{-2\xi_1}+\rme^{-2\xi_2}+\rme^{-2\xi_3}) + \rme^{-\varphi}\triangle\varphi = 0.
\end{equation}
Due to \eqref{5dgen: 1} one of the four equations in \eqref{5dgen: bi} and \eqref{5dgen: 3} is redundant. We can therefore use \eqref{5dgen: 1} to eliminate $\xi_1$ and work with a system of two coupled PDEs for $\xi_2=\xi$ and $\xi_3=\chi$ in addition to the following equation for $\varphi$ obtained from \eqref{5dgen: extra}
\begin{equation}\label{5dgen: fin}
    \partial^2_\rho \rme^\varphi - 4(\rme^{-\xi}+\rme^{-\chi})\partial_\rho \rme^\varphi + 8\rme^\varphi(\rme^{-2\xi}+\rme^{-2\chi}+\rme^{-\xi-\chi}) + \triangle\varphi =0\,.
\end{equation}
Therefore we conclude that by solving the equations for $\varphi$, $\chi$, and $\xi$ in \eqref{5dgen: fin} and \eqref{5dgen: bi} we can find the most general supersymmetric background of the form in \eqref{eq:5dstuansatz} in the five-dimensional STU supergravity model. 

\subsubsection{IR analysis}
\label{subsubsec:5dIRstu}
We begin by deriving the AdS$_3\times \Sigma_{\mathfrak{g}}$ IR solutions of \cite{Benini:2013cda} in our notation. To this end we take
\begin{equation}\label{5dgen: irans}
        \varphi = -\log\varphi_0 + 2\log\rho +2 g(x,y)\,, \hspace{0.5cm}
 \xi_a = -\log\xi_{a,0} + \log\rho\,.
\end{equation}
Here $\varphi_0$ and $\xi_{a,0}$ are constants and the function $g$ has to obey the Liouville equation \eqref{5DLiouville} and thus determines a metric on $\Sigma_{\mathfrak{g}}$ of constant curvature $\kappa$. The field strengths of the three gauge are proportional to the volume form on $\Sigma_{\mathfrak{g}}$
\begin{equation}
    F^a = -n_a \rme^{2g(x,y)} \rmd{x}\wedge \rmd{y}\,,
\end{equation}
where, as described in \cite{Benini:2013cda}, $n_a$ are constant magnetic flux parameters which need to be quantized in terms of the genus $\mathfrak{g}$.

With this at hand we find that equations \eqref{5dgen: 1}, \eqref{5dgen: 2}, and \eqref{5dgen: 3} reduce to the following algebraic constraints
\begin{equation}
\begin{split}
     &\xi_{1,0} + \xi_{2,0} + \xi_{3,0}=1\,,\\
    &n_1 + n_2 + n_3 = \kappa\,,\\
    & 2\xi_{a,0} - n_a \frac{\varphi_0}{\xi_{a,0}}=1, \qquad a =1,2,3 \,.
    \end{split}
\end{equation}
For generic values of the parameters $n_a$ these equations can be solved to find
\begin{equation}\label{5dgen: ir}
\begin{split}
     \varphi_0 &=-\frac{\Pi}{\Theta^2}\,, \hspace{3cm}  \xi_{1,0} = \frac{n_1(n_1-n_2-n_3)}{\Theta}\,,\\
     \xi_{2,0} &=\frac{n_2(-n_1+n_2-n_3)}{\Theta}\,, \qquad \xi_{3,0} =\frac{n_3(-n_1-n_2+n_3)}{\Theta}\,,
\end{split}
\end{equation}
where, as in \cite{Benini:2013cda}, we have defined
\begin{equation}\begin{split}
    \Pi &= (-n_1+n_2+n_3)(n_1-n_2+n_3)(n_1+n_2-n_3)\,,\\
    \Theta &=n^2_1 + n^2_2+n^2_3 - 2(n_1n_2+n_2n_3+n_3n_1)\,.
\end{split}
\end{equation}
We have thus recovered the supersymmetric AdS$_3$ solutions discussed in \cite{Benini:2013cda}. It is important to emphasize that in order for these solutions to be physically acceptable the magnetic fluxes $n_a$ have to obey certain positivity constraints. These arise since, due to \eqref{5dgen: irans}, the constants $\varphi_0$ and $\xi_{I,0}$ have to be positive. These constraints were thoroughly analyzed in \cite{Benini:2013cda} and from now on we assume that we work with values of $n_a$ for which they are obeyed. 

To study whether the constant curvature metric on $\Sigma_{\mathfrak{g}}$ can be deformed as we move away from the IR region we proceed as in the previous sections. Namely, we expand equations \eqref{5dgen: bi} and \eqref{5dgen: fin} around the IR AdS$_3\times \Sigma_{\mathfrak{g}}$ solution described above in terms of eigenmodes of the constant curvature Laplacian on $\Sigma_{\mathfrak{g}}$. The radial evolution of every such eigenmode, with eigenvalue $\mu_n$, can be obtained by linearizing the equations in \eqref{5dgen: bi} and \eqref{5dgen: fin} to find
\begin{equation}\label{eq:5dstulinIR}
    \begin{split}
        &\rho^2 \partial^2_\rho \tilde{\xi}_n + 4\xi_0 \rho\partial_\rho \tilde{\xi}_n + (1-2\xi_0)\rho\partial_\rho \tilde{\varphi}_n -\varphi_0 \mu_n \tilde{\xi}_n =0\,,\\
       & \rho^2 \partial^2_\rho \tilde{\chi}_n + 4\chi_0 \rho\partial_\rho \tilde{\chi}_n + (1-2\chi_0)\rho\partial_\rho \tilde{\varphi}_n -\varphi_0 \mu_n \tilde{\chi}_n =0\,,\\
        &\rho^2 \partial^2_\rho \tilde{\varphi}_n + 4(1-\xi_0-\chi_0)\rho\partial_\rho\tilde{\varphi}_n  + 8 (\xi_0\tilde{\xi}_n(1-2\xi_0-\chi_0) + \chi_0\tilde{\chi}_n(1-\xi_0 -2\chi_0))\\&\hspace{9.5cm}-(\mu_n-2\kappa )\varphi_0 \tilde{\varphi}_n=0\,.
    \end{split}
\end{equation}
The regular solutions of these equations in the IR region at $\rho \to 0$  take the form
\begin{equation}\label{5dpertans}
     \begin{pmatrix}
   \tilde{\xi}_n\\\tilde{\chi}_n\\\tilde{\varphi}_n\end{pmatrix}
 = \rho^{\gamma}    \begin{pmatrix}A\\B\\C\end{pmatrix}\,,
\end{equation}
where $(A,B,C)$ are real integration constants and $\gamma$ should be a positive real number. Note that the value of the constant $\gamma$ depends on $n$, i.e. on the choice of eigenmode. Plugging \eqref{5dpertans} in \eqref{eq:5dstulinIR} we obtain the following algebraic equation
\begin{equation}\label{5dirmat}
    M  \begin{pmatrix}A\\B\\C\end{pmatrix} =0\,, 
\end{equation}
where the matrix $M$ is given by
\begin{equation}
     M =  \begin{pmatrix}\gamma^2 +\gamma(4 \xi_0-1) - \mu_n \varphi_0&0 &\gamma(1-2\xi_0)\\&&\\0 &\gamma^2 + \gamma(4 \chi_0-1) - \mu_n \varphi_0 & \gamma(1-2\chi_0)\\&&\\8\xi_0(1-2\xi_0-\chi_0)& 8\chi_0(1-\xi_0-2\chi_0)&\gamma^2 + \gamma(3-4\xi_0-4\chi_0)\\& & +\varphi_0(2\kappa -\mu_n)  \end{pmatrix}\,.\notag
\end{equation}
In order to show that for every eigenvalue $\mu_n$ there is a regular solution of the IR linearization problem as in \eqref{5dpertans} we need to ensure that the algebraic equation ${\rm det}M=0$ always have a positive root. First we note that when the magnetic fluxes are equal, i.e. $n_1=n_2=n_3$, we find $\varphi_0 = \chi_0 = \xi_0 = \frac13$ and recover the solution of the minimal gauged supergravity with the value of $\gamma$ in \eqref{eq:5dmingamman}. For general values of the flux parameters we have to solve a sixth order polynomial equation of the form
\begin{equation}\label{eq:detM5d}
    \det (M) = \gamma^6 + \gamma^5 +... - \varphi^3_0 \mu^2_n(\mu_n-2\kappa) =0\,.
\end{equation}
We do not have a closed form expression for the roots of this polynomial equations. Nevertheless the structure of the terms we presented explicitly in \eqref{eq:detM5d} allows us to deduce important information. In particular, the positivity constraints on $\varphi_0$ and $\mu_n$ and the form of the constant term in \eqref{eq:detM5d} imply that the product of the roots is negative. Since we have a sixth order polynomial this in turn implies that at least one of the roots is positive for almost all values of $\kappa$ and $\mu_n$. The only exceptions to this argument are the zero modes $\mu_0 = 0$ for all values of $\kappa$ and the mode $\mu_1 = 2$ when $\kappa =1$. In these two cases we have checked numerically that  for general values of $n_a$ allowed by the positivity constraints in \cite{Benini:2013cda} there are three positive roots of the equation in \eqref{eq:detM5d}. We have also analyzed the polynomial equation \eqref{eq:detM5d} for more general values of $\mu_n$ and $\kappa$. We find that for values of $n_a$ that obey the positivity constraints in \cite{Benini:2013cda}, there are three positive real roots. Moreover the corresponding solutions of the form \eqref{5dpertans} have $C \neq 0$. This is important to ensure that the metric on the Riemann surface is indeed perturbed since, due to \eqref{eq:5dvarphidefstu}, this perturbation is encoded in $\varphi$. The fact that there are three positive roots in general means that not only the metric perturbations but also the scalar perturbations are washed out as we approach the IR region. A special situation which necessitates a separate analysis arises when two of the fluxes re equal. Then the polynomial in \eqref{eq:detM5d} factorizes into two polynomials of degree 4 and 2, respectively. The degree 2 polynomial has two positive roots corresponding to an eigenvector with $C=0$. These special modes leave the scalar fields and the metric function $f$ unperturbed and correspond to the marginal deformations discussed around Equation (21) in \cite{Maldacena:2000mw}. The degree 4 polynomial however, still contains a positive root corresponding to an eigenvector with $C\neq 0$ which ensures that we can indeed perturb the metric away from the constant curvature one as we move away from the IR region.

\subsubsection{UV analysis}

In the UV region we encounter a familiar predicament. Namely, we can solve equation \eqref{5dgen: bi} and \eqref{5dgen: fin} order by order in the $\rho \to \infty$ limit and then find that the metric on the Riemann surface can be arbitrary. We find the following UV expansion for the three unknown functions that control the BPS black string solutions
\begin{equation}\begin{split}
\varphi =&\,3 \log\rho + \varphi_0 + \varphi_{1,1}\rho^{-1} \log\rho + \varphi_1 \rho^{-1}+  \varphi_{2,2}\rho^{-2}(\log\rho)^2\\
&\hspace{6.5cm}  + \varphi_{2,1}\rho^{-2}\log\rho+ \varphi_2 \rho^{-2} + \ldots \,,\\
\xi =&\, \log\rho + \log2 +\xi_{1,1}\rho^{-1} \log\rho + \xi_1 \rho^{-1} + \xi_{2,2}\rho^{-2}(\log\rho)^2\\
&\hspace{6.5cm}+ \xi_{2,1} \rho^{-2} \log\rho +\xi_2 \rho^{-2} +\ldots\,,\\
\chi =&\,-2 \log\rho + \log(2) +\chi_{1,1}\rho^{-1} \log\rho + \chi_1 \rho^{-1} + \chi_{2,2}\rho^{-2}(\log\rho)^2\\
&\hspace{6.5cm}+ \chi_{2,1} \rho^{-2} \log\rho +\chi_2 \rho^{-2} + \ldots\,,
\end{split}\end{equation}
where all of the coefficients are functions of the two coordinates on the Riemann surface. As in the UV expansion of the BPS equation in the minimal five-dimensional supergravity discussed in Section~\ref{subsubsec:5dminUV} we find two undetermined metric functions, $\varphi_0$, $\varphi_1$. In addition to that we have four other undetermined functions, $\xi_1$, $\chi_1$, $\xi_{1,1}$, and $\chi_{1,1}$, which are due to the presence of the additional matter fields in the STU model.  The rest of the coefficients in the UV expansion are determined in terms of these six unknown functions. For instance for the first few coefficients in the expansion of $\xi$ we find
\begin{equation}
    \begin{split}
    \xi_{2,2} &= -\frac12 \xi^2_{1,1}\,,\\
    \xi_2 &= -\frac12\xi^2_1 + \varphi_1\xi_{1,1}- \varphi_{1,1}\xi_{1,1}+2\xi_{1,1}^2+2 \rme^{-\varphi_0}(\partial^2_x + \partial^2_y)\xi_{1,1} +2 \xi_{2,1}\,,\\
    \xi_{2,1} &= \varphi_{1,1}\xi_{1,1}-\xi_1\xi_{1,1}-2\xi_{1,1}^2-e^{-\varphi_0}(\partial^2_x + \partial^2_y)\xi_{2,1}\,.
\end{split}
\end{equation}
while for the first two coefficients in $\varphi$ we have
\begin{equation}\begin{split}
    \varphi_{2,2}&= -\Big(\chi^2_{1,1} + \xi^2_{1,1} + \frac12\varphi^2_{1,1}-\phi_{1,1}(\chi_{1,1}+\xi_{1,1})+\chi_{1,1}\xi_{1,1}\Big)\,,\\
        \varphi_{1,1} &= \rme^{-\varphi_0}(\partial^2_x + \partial^2_y)\varphi_0\,.
    \end{split}
\end{equation}
Similar expressions can be obtained for the low order coefficients in the expansion of $\chi$ as well. 

As in the previous examples we studied we thus conclude that the metric on $\Sigma_{\mathfrak{g}}$ is arbitrary in the UV region and should approach the constant curvature one in the IR region of these regular five-dimensional solution.

 \subsection{Four-dimensional STU model}

We now proceed to study a large generalization of the black hole solution discussed in Section~\ref{4DUniversal}. To this end we focus on the STU model of four-dimensional $\mathcal{N}=2$ supergravity which arises as a particular consistent truncation of of eleven-dimensional supergravity on $S^7$. This model can also be constructed by adding three $\mathcal{N}=2$ vector multiplets to the gravity multiplet of the minimal supergravity theory. Each of the vector multiplets contains an $\U(1)$ gauge field and two complex scalars which can thought of as a scalar and a pseudoscalar. For the magnetic black hole solutions of interest here it can be shown that the pseudoscalars can be consistently set to zero. The bosonic Lagrangian of this model is given by\footnote{We follow closely \cite{Benini:2015eyy} in our presentation, see also \cite{Cacciatori:2009iz}.}
\begin{equation}\label{eq:4dSTULag}
    \mathcal{L} = R - \frac12 (\partial \vec{\phi})^2 - \frac14 \sum_{a=1}^4 \rme^{\vec{\mathfrak{a}}_a\cdot\vec{\phi}}F^2_a - V(\phi)\,,
\end{equation}
where the potential is
\begin{equation}
    V = -2(\cosh{\phi_{12}}+ \cosh{\phi_{13}} + \cosh{\phi_{14}})\,,
\end{equation}
and we have defined
\begin{equation}
\begin{split}
     &\vec{\mathfrak{a}}_1 = (1,1,1), \hspace{0.5cm}  \vec{\mathfrak{a}}_2 = (1,-1,-1), \hspace{0.5cm}  \vec{\mathfrak{a}}_3 = (-1,1,-1) \\ & \vec{\mathfrak{a}}_4 = (-1,-1,1), \hspace{0.5 cm}\vec{\phi} = (\phi_{12},\phi_{13}, \phi_{14})\,.
\end{split}
\end{equation}
We have not included the $F \wedge F$ terms in the Lagrangian \eqref{eq:4dSTULag} since they will not play any role for the magnetic black hole solutions of interest here.

The STU model arises also as a consistent truncation of the $\mathcal{N}=8$ $\SO(8)$ gauged supergravity. This embedding proves useful when studying the supersymmetry variations of the theory. Im the $\mathcal{N}=8$ theory the fermions consist of the gravitini $\psi^I_\mu$ and the spin-1/2 fields $\chi^{IJK}$ where $I$, $J$ and $K$ are $\SU(8)$ indices. In the $\mathcal{N}=2$ STU model truncation the index $I$ should be thought of as corresponding to the pair $(a, i)$ where $a = 1, \ldots , 4$ as in \eqref{eq:4dSTULag} and $i = 1, 2$. With this notation the supersymmetry variations of the gravitini are given by 
\begin{equation}\label{eq:susyvar324dstu}
\begin{split}
\delta\psi^{a i}_\mu = &\nabla_\mu \epsilon^{a i}-\frac12 \Omega_{ab}A^b_{\mu}\epsilon^{ij}\epsilon^{b j}+ \frac18\sum_b \rme^{-\vec{\mathfrak{a}}_b\cdot\vec{\phi}/2}\gamma_\mu \epsilon^{a i} \\
&\hspace{3.2cm}+ \frac18\sum_b \Omega_{ab}\rme^{\vec{\mathfrak{a}}_b\cdot\vec{\phi}/2}F^b_{\nu\lambda}\gamma^{\nu\lambda}\gamma_\mu \epsilon^{ij}\epsilon^{a j}\,,
\end{split}
\end{equation}
where $\Omega_{ab}$ is the matrix
\begin{equation}
    \Omega = \frac12\begin{pmatrix}
1&1&1&1\\
1&1&-1&-1\\
1&-1&1&-1\\
1&-1&-1&1
\end{pmatrix}\,.
\end{equation}
For the spin-$1/2$ variations, one finds $\delta \chi^{a i\;b j\;c k} = \delta \chi^{a\;c k}\delta^{a \beta}\epsilon^{ij} + \delta \chi^{b\;a i}\delta^{b c}\epsilon^{jk} + \delta \chi^{c\;b j}\delta^{c a}\epsilon^{ki}$ where 
\begin{equation}\label{eq:susyvar124dstu}
\begin{split}
    \sqrt{2}\; \delta \chi^{a\;b i} = & -\gamma^\mu \partial_\mu \phi_{ab}\epsilon^{ij}\epsilon^{b j} -\sum_{cd} \Sigma_{abc}\Omega_{cd}e^{-\vec{\mathfrak{a}}_d\cdot\vec{\phi}/2}\epsilon^{ij}\epsilon^{b j} \\
    &\hspace{4.2cm}+ \sum_d \Omega_{ad}e^{\vec{\mathfrak{a}}_d\cdot\vec{\phi}/2}F^d_{\mu\nu}\gamma^{\mu\nu}\epsilon^{b i}\,.
\end{split}
\end{equation}
Note that there is no sum over the repeated index $b$ above. The tensor $\Sigma_{abc}$ is defined as
\begin{equation}
 \Sigma_{abc}=
\begin{cases}
|\epsilon_{abc}| &\text{for } a, b, c \neq 1\,,\\
\delta_{bc}  &\text{for } a = 1\,,\\
\delta_{ac} &\text{for } b=1\,,\\
0                     &\text{otherwise}\,.
\end{cases}   
\end{equation}
%

\subsubsection{BPS constraints}

to construct the black hole solutions of interest here we consider the following Ansatz, for $a=1,2,3,4$, 
\begin{equation}\label{eq:4dstuansatz}
    \begin{split}
    \rmd{s}^2 &= -\rme^{2f(r,x,y)}\rmd{t}^2 + \rme^{2h(r,x,y)}\rmd{r}^2+\rme^{2g(r,x,y)}(\rmd{x}^2+\rmd{y}^2)\,,\\
    A^a  &= A^a_x(r,x,y) \rmd{x} + A^a_y (r,x,y)\rmd{y}\,,\\
     \rme^{-\vec{\mathfrak{a}}_a \cdot \vec{\phi}/2}&= X_a(r,x,y)\,.
    \end{split}
\end{equation}
To recover the solutions in the minimal supergravity discussed in Section~\ref{4DUniversal} we need to set the four gauge fields equal to each other and freeze the scalar fields by setting $X_a=1$. The spinor generating the supersymmetry of this background again takes the simple form $\epsilon^{a i} = \rme^{f/2}\epsilon^{a i}_0$, where $\epsilon^{a i}_0$ are constant spinor parameters. Since the solutions preserve only two real supercharges we take only the $a=1$ components of the spinor to be non-vanishing. In addition, we impose the projectors $\gamma_{\hat{r}}\epsilon^i_0=\epsilon^i_0$ and $\gamma_{\hat{x}\hat{y}}\epsilon^i_0 = -\varepsilon^{ij} \epsilon^j_0$. By imposing that the spin-1/2 supersymmetry variations vanish we then can derive the following constraints on the bosonic fields
\begin{eqnarray}\label{4dgen: dil1}
    \rme^{-h}\partial_r \phi_{1a} + \Omega_{ab}X_b +\rme^{-2g}\Omega_{ab}X^{-1}_bF^b_{\mu\nu}=0\,,\\
\label{4dgen: dil2}\rme^h (\partial_x + i \partial_y)\phi_{1a} + i\Omega_{ab}X^{-1}_b(F^b_{rx}+iF^b_{ry})=0\,.
\end{eqnarray}
From the $t$-component of the gravitino variations we find the relations
\begin{eqnarray}\label{4dgen: grav1}
    \rme^{-h}\partial_r f + \frac12  \Omega_{1b}X_b + \frac12 \rme^{-2g}\Omega_{1b}X^{-1}_bF^b_{xy}=0\,,\\
\label{4dgen: grav2}\rme^h(\partial_x + i \partial_y)f + \frac{i}{2}\Omega_{1b}X^{-1}_b(F^b_{rx}+ iF^b_{ry})=0\,. 
\end{eqnarray}
The expressions above determine the sum of the field strengths which in turn can be used to simplify the rest of the supersymmetry variations. The result for the constraints derived from the other components of the gravitino variations reads
\begin{eqnarray}\label{4dgen: con1}
(\partial_x + i \partial_y)(f+h) &=& 0\,,\\
\label{4dgen: con2}\partial_r (f+g)+ \rme^h \Omega_{1b}X_b &=&0\,,\\
2\partial_y(f+g)+\sum_b A^b_x &=& 0\,,\\
2\partial_x(f+g)-\sum_b A^b_y &=&0\,.
\end{eqnarray}
These relations should be compatible with the ones in \eqref{4dgen: grav1} and \eqref{4dgen: grav2} which leads to the following alternative expressions for the field strengths
\begin{eqnarray}\label{4dgen: con3}
    \sum_b F^b_{xy} - 2\triangle(f+g)&=&0\,,\\
    \label{4dgen: con4}i\sum_b (F^b_{rx}+ i F^b_{ry}) + 2\partial_r(\partial_x + i \partial_y)(f+g)&=&0\,.
\end{eqnarray}
We have so far derived all constraints on the bosonic fields imposed by the vanishing of the supersymmetry variations in \eqref{eq:susyvar324dstu} and \eqref{eq:susyvar124dstu}. To simplify these BPS equations further we use \eqref{4dgen: con1} to change coordinates by $\rmd{\rho}=- \rme^{f+h}\rmd{r}$. We can then rewrite \eqref{4dgen: dil1}-\eqref{4dgen: con4} in a simpler way upon defining\footnote{Note that the definition of $\varphi$ here differs by a factor 4 from the one used in Section~\ref{4DUniversal}.} $\varphi =2f+2g$ and $\xi_a = f + \vec{\mathfrak{a}}_a \cdot \vec{\phi}/2$, for $a=1,2,3,4$.  This yields the following system of equations
\begin{equation}
\label{4dgen: 1}\partial_\rho \varphi = \sum_a \rme^{-\xi_a}\,,
\end{equation}
\begin{equation}
  \label{4dgen: 2}\sum_aF^a_{xy} - \triangle\varphi=0  \,,
\end{equation}
\begin{equation}
    \label{4dgen: 4}\partial_\rho \rme^{-\xi_a} + \rme^{-2\xi_a} + \rme^{ - \varphi}F^a_{xy}=0\,, \hspace{0.8cm}a=1,2,3,4
\end{equation}
\begin{equation}
    \label{4dgen: 5}(\partial_x + i\partial_y)\rme^{-\xi_a} = i (F^a_{rx}+iF^a_{ry})\rme^{-f-h}\,. \hspace{0.8cm}a=1,2,3,4.
\end{equation}
To simplify these equations even further we can eliminate the field strengths by imposing the Bianchi identity
\begin{equation}
    \partial_r F^a_{xy} + \partial_x F^a_{yr} + \partial_y F^a_{rx}=0\,.
\end{equation}
This allows us combine to  \eqref{4dgen: 4} and \eqref{4dgen: 5} to get, for $a=1,2,3,4$, 
\begin{equation}\label{4dgen: bi}
    \partial_\rho \lbrack \rme^\varphi (\partial_\rho \rme^{-\xi_a}+ \rme^{-2\xi_a}) \rbrack + \triangle\rme^{-\xi_a}=0\,,
\end{equation}
together with 
\begin{equation}\label{4dgen: last}
    \sum_a \lbrack \partial_\rho \rme^{-\xi_a} + \rme^{-2\xi_a}\rbrack + \rme^{ - \varphi} \triangle\varphi=0\,,
\end{equation}
obtained from \eqref{4dgen: 2} and \eqref{4dgen: 4}.

The equations in \eqref{4dgen: bi} and \eqref{4dgen: last} encode the conditions imposed by supersymmetry on the bosonic backgrounds of the form \eqref{eq:4dstuansatz}. We will use them to understand the behavior of perturbations away from the constant curvature black hole solutions studied in \cite{Cacciatori:2009iz,Benini:2015eyy}. We note that not all of these equations are independent since the sum of the equations in \eqref{4dgen: bi} gives us the radial derivative of \eqref{4dgen: last}. 

\subsubsection{IR analysis}

In the IR region the only regular solutions are the AdS$_2$ for which the metric on $\Sigma_{\mathfrak{g}}$ is constant, see \cite{Gauntlett:2001qs,Cacciatori:2009iz,Benini:2015eyy}. In our notation these backgrounds take the form
\begin{equation}\label{eq:ads2stusoln4d}
  \varphi = -\log\varphi_0+2\log\rho+ 2g(x,y), \hspace{0.5cm}\xi_a = -\log\xi_{a,0}+ \log\rho\,,
\end{equation}
where $\varphi_0$ and $\xi_{a,0}$ are constants. The function $g(x,y)$ solves the Liouville equation in \eqref{4DLiouville} and determines a metric of constant curvature $\kappa$ on the Riemann surface. The field strengths are then proportional to the volume form on the Riemann surface
 \begin{equation}
     F^a = -n_a \rme^{2g(x,y)} \rmd x\wedge  \rmd y\,,
 \end{equation}
where $n_a$ are the magnetic flux parameters that need to be quantized as discussed in \cite{Benini:2015eyy}. The BPS equations for these AdS$_2\times \Sigma_{\mathfrak{g}}$ solution then reduce to the following algebraic relations
\begin{equation}\label{eq:algebBPSads2}
    2\kappa = \sum_a n_a\,, \qquad     2 = \sum_a \xi_{a,0}\,, \qquad     \xi_{a,0} - n_a \frac{\varphi_0}{\xi_{a,0}}=1 \qquad \text{for }a=1,2,3,4
\end{equation}
To solve these equations we proceed as in \cite{Benini:2015eyy} and define 
\begin{equation}
    \Pi = \frac18 (n_1 + n_2 - n_3 - n_4)(n_1 - n_2 + n_3 - n_4)(n_1 - n_2 - n_3 + n_4)\,,
\end{equation}
together with
\begin{equation}
    \Theta = (\mathcal{F}_2)^2 - 4n_1n_2n_3n_4, \hspace{0.5cm}\mathcal{F}_2 = \frac14\Big(\sum_a n_a\Big)^2 - \frac12\sum_a n^2_a\,.
\end{equation}
The solutions of the equations in \eqref{eq:algebBPSads2} can then be written as
\begin{equation}
    \varphi_0 = \frac{\Pi}{\Theta}, \hspace{0.5cm} \xi_{a,0} = \frac12 \pm \frac{\mathcal{F}_2 + n_a(n_1 + n_2+n_3+n_4 - 2n_a)}{2\sqrt{\Theta}}\,.
\end{equation}
To ensure that these AdS$_2\times\Sigma_{\mathfrak{g}}$ solutions are well-defined the constants in \eqref{eq:ads2stusoln4d} have to obey certain positivity constraints. This in turn leads to constraints on the values of the magnetic flux parameters $n_a$ which are analyzed in \cite{Benini:2015eyy}. From now on we assume that we always take the parameters $n_a$ to take values in the allowed regions of parameter space.

Our goal now is to study perturbations around this constant curvature solution away from the IR region and show that the metric on $\Sigma_{\mathfrak{g}}$ can deviate from the constant curvature one as one moves towards the UV region. To this end we expand the functions in \eqref{4dgen: bi} and \eqref{4dgen: last} in eigenmodes of the Laplacian for the constant curvature metric on $\Sigma_{\mathfrak{g}}$ and study the behavior of these linearized perturbations. Since the equations in \eqref{4dgen: bi} and \eqref{4dgen: last} are not independent it is most convenient to work only with three out of the four equations in \eqref{4dgen: bi} which yield the following linear equations for the eigenmodes $\tilde{\xi}_{a,n}$ and $\tilde{\varphi}_n$
\begin{equation}
      \rho^2 \partial^2_\rho \tilde{\xi}_{a,n} +  2 \rho \xi_{a,0}\partial_\rho \tilde{\xi}_{a,n} + (1-\xi_{a,0})\rho\partial_\rho \tilde{\varphi}_n - \varphi_0 \mu_n \tilde{\xi}_{a,n}=0\,.
\end{equation}
The regular solutions of this system of equations is determined by the following vector of solutions\footnote{The solution for the modes $\xi_{4,n}$ is not independent and is obtained by exploiting the relations between the equations in \eqref{4dgen: bi} and \eqref{4dgen: last}.}
 \begin{equation}\label{eq:4dstugammaeqn}
    \begin{pmatrix}\tilde{\xi}_{1,n} & \tilde{\xi}_{2,n} & \tilde{\xi}_{3,n} & \tilde{\varphi}_n \end{pmatrix}^\intercal = \rho^\gamma \vec{v}.
    \end{equation}
where the real constant $\gamma$ and the constant four-vector $\vec{v}$ depend on the choice of eigenmode $n$. To find this solution we need to use \eqref{4dgen: 1}, \eqref{4dgen: bi}, and \eqref{4dgen: last} along with the linearized expansion described above. This results in a matrix equation of the form $M\cdot\vec{v} =0$ where $M$ is a $4\times 4$ matrix whose explicit form is too unwieldy to present here. To analyze this we proceed as in Section~\ref{subsubsec:5dIRstu} and find the  constants $\gamma$ by solving the algebraic equations $\det M =0$. This results in an eighth degree polynomial equation for $\gamma$ that is too complicated to analyze analytically and we have resorted to a numerical analysis. We have checked explicitly that for many choices of the magnetic flux parameters $n_a$ the algebraic equation $\det M =0$ has four positive roots for the constants $\gamma$. When we use these numerical solutions in \eqref{eq:4dstugammaeqn} we find that they correspond to perturbations of the metric and the bosonic fields in the black hole solution \eqref{eq:4dstuansatz} which deviate from the constant curvature solution but are small and ultimately washed out near the AdS$_2$ region in the IR. Therefore we once again observe the characteristic holographic uniformization behavior.

\subsubsection{UV analysis}

To complete our arguments in favor of the holographic uniformization behavior of the BPS equations we analyze the solutions of the equations in \eqref{4dgen: bi},and \eqref{4dgen: last} in the UV region $\rho \to \infty$. The results of this analysis are familiar by now. We find that the metric on the Riemann surface is indeed allowed to be arbitrary in this UV region. To be more explicit we can expand $\varphi$, $\xi=\xi_1$, $\psi=\xi_2$, and $\chi=\xi_3$ in the following form
\begin{equation}
    \begin{split}
        \varphi &= 4\log\rho+ \sum_n \varphi_n(x,y) \rho^{-n}\,, \qquad \xi =  \log\rho +  \sum_n \xi_n(x,y) \rho^{-n}\,, \\
        \psi &=  \log\rho +  \sum_n \psi_n(x,y) \rho^{-n}\,, \qquad \chi =  \log\rho +  \sum_n \chi_n(x,y) \rho^{-n}\,.
    \end{split}
\end{equation}
The functions $\varphi_0$, $\varphi_1$ as well as $\xi_1, \chi_1, \psi_1$ and $\xi_2, \psi_2, \chi_2$ in the expression above are left undetermined by the BPS equations. The freedom to choose these functions arbitrarily reflects the choice of metric on $\Sigma_{\mathfrak{g}}$ in the UV region.

The higher order coefficients in the UV expansion are completely determined in terms of these functions. For instance, we at second order we find the relation
\begin{equation}\begin{split}
    \varphi_2 = &- \frac12 \rme^{-\varphi_0}(\partial^2_x+\partial^2_y)\varphi_0 - (\xi^2_1+ \psi^2_1+\chi^2_1 + \xi_1\psi_1 +\xi_1\chi_1 + \chi_1\psi_1)\\&\hspace{7cm} -\frac12 \varphi^2_1 + \varphi_1(\xi_1 + \psi_1 + \chi_1)\end{split}\,.
\end{equation}
At third order we find
\begin{equation}
    \xi_3 = \frac13 \xi^3_1 - \frac14 \varphi_1 \xi^2_1 - \frac12 \varphi_1 \xi_2 - \frac12\rme^{-\varphi_0}(\partial^2_x+\partial^2_y)\xi_1\,,
\end{equation}
with similar expressions for $\chi_3$ and $\psi_3$. Also at this order, $\varphi_3$ is fixed in terms of the lower order coefficients as
\begin{equation}
    \begin{split}
        \varphi_3 = \; &\xi_3 + \chi_3 + \psi_3 + \frac16 (\xi^3_1 + \psi^3_1 + \chi^3_1 )-\xi_1\xi_2-\psi_1\psi_2-\chi_1\chi_2\\
        &+\frac12(\xi_1 + \chi_1 + \psi_1 - \varphi_1)(\xi^2_1 + \chi^2_1+\psi^2_1-2\xi_2-2\chi_2-2\psi_2 + 4\varphi_2)\\
        &-\frac{1}{6}\varphi_1\lbrack\varphi^2_1+2\xi_1\psi_1 + 2\psi_1\chi_1+2\chi_1\xi_1 + 2\varphi_2 - (\xi_2+\psi_2+\chi_2)\rbrack\\
        &-\frac{1}{4}\varphi_1(\xi^2_1+\psi^2_1+\chi^2_1)+ \frac13\varphi_1(\xi_1+\psi_1+\chi_1)\\
        &+\frac16\rme^{-\varphi_0}(\partial^2_x+\partial^2_y)(\xi_1+\chi_1+\psi_1 - \varphi_1)\,.
    \end{split}
\end{equation}
%

\section{Conclusion}
\label{sec:conclusions}

In this paper we showed that there are large families of supersymmetric asymptotically locally AdS black brane solutions with a smooth Riemann surface horizon geometry in gauged supergravity. At asymptotic infinity, i.e. the UV region of the geometry, the metric on the Riemann surface can be chosen freely. However, the supergravity BPS equations result in non-linear PDEs which uniformize the metric on the Riemann surface such that in the near horizon region it is fixed to the constant curvature metric. These results generalize and extend the holographic uniformization discussed in \cite{Anderson:2011cz}. They also bear resemblance to the well-known attractor mechanism for asymptotically flat black holes with the notable difference that in our examples the moduli at asymptotic infinity are not scalar fields arising from the internal dimensions of string or M-theory.

Our work presents a number of open questions and possible generalizations. Here we list some of them.

\begin{itemize} 

\item It is desirable to perform a more mathematically rigorous analysis of the PDEs resulting from the supergravity BPS equations that we derive in this paper.  This should proceed using similar methods as the ones employed in \cite{Anderson:2011cz} and should lead to a global existence proof of the smooth uniformization solutions.

\item All gauged supergravity solutions studied in this paper are in theories which arise as consistent truncations from ten- or eleven-dimensional supergravity. It is therefore possible to uplift the solutions we constructed above to string or M-theory and it will be interesting to do so explicitly. This applies especially to the solutions discussed in Section~\ref{sec:universal} which admit various different embeddings in higher dimensions, see for example \cite{Bobev:2017uzs} for a detailed discussion and a list of references.

\item We have opted to study several specific gauged supergravity theories which are particularly simple and explicit and in addition can be embedded in string theory. Our general results should apply more broadly and it should be possible to find black brane solutions of more general matter matter coupled supergravity theories which exhibit similar attractor mechanism for the horizon metric.

\item There is a large body of literature on studying the constraints imposed by supersymmetry, and the resulting differential equations, for solutions of various supergravity theories in different dimensions. We have not used these results to derive the BPS equations and the resulting PDEs analyzed in this paper. Nevertheless, it should be possible to rephrase our results in these more general terms. In particular the solutions discussed in Section~\ref{sec:universal} should fit into the classification results of \cite{Caldarelli:2003pb}, \cite{Gauntlett:2003fk}, \cite{Cariglia:2004kk}, and \cite{MacConamhna:2004fb} for four-, five-, six-, and seven-dimensional minimal supergravity, respectively.

\item An important assumption in our work is that the supergravity solutions we study are static and the Riemann surface describing the horizon has a smooth metric. It should be possible to relax both assumptions, for example by generalizing the four-dimensional solutions with angular momentum discussed in \cite{Hristov:2018spe,Hristov:2019mqp} and by studying the solutions of  \cite{Bobev:2019ore} with a punctured Riemann surface away from the near horizon region.

\item Based on the physics of topologically twisted SCFTs on compact manifolds, one should expect that the holographic uniformization principle is not limited to Riemann surfaces. It would be very interesting to understand this vast generalization in particular in the context of hyperbolic three-manifolds \cite{BBGP}, as well as four-manifolds where some initial studies were performed in \cite{Fluder:2017nww,Fluder:2017hrb}. In this context holographic uniformization and its generalizations offer another example of the close relation between the physics of RG flows and the mathematics of geometric flows, similar in spirit to Ricci flow \cite{Friedan:1980jf} \cite{Perelman:2006un}.

\item The uniformization nature of the black brane solutions we found ensures that the metric of the horizon is the one with constant curvature and thus the horizon area and black hole entropy do not depend on continuous moduli. This attractor mechanism is an important feature of our solutions. Another important quantity for black hole thermodynamics is the regularized on-shell action. For the four-dimensional constant curvature black hole in Section~\ref{subsubsec:cc4dBH} it was shown in \cite{Azzurli:2017kxo}, see also \cite{Cabo-Bizet:2017xdr}, that the regularized on-shell action is equal to the black hole entropy. It will be very interesting to study whether this is true more generally for the black brane solutions studied here. The results in \cite{BenettiGenolini:2019jdz}, and their generalizations to higher dimensions, should prove useful in establishing this question.

\item In establishing the attractor behavior for the metric moduli near the black brane horizon we assumed that the solution is supersymmetric. This is important for two reasons. First, the BPS equations of supergravity are presumably technically simpler to analyze then the full equations of motion. Second, the supersymmetry of the near horizon AdS region ensures that the black hole solution is stable against small perturbations. It will be very interesting to understand whether the general lessons from our results are true for general non-supersymmetric charged black branes in AdS. Perhaps a natural starting point to address this questions is to study non-supersymmetric extremal black brane solutions similar to the ones studied in \cite{DHoker:2009mmn,Almuhairi:2010rb,Almuhairi:2011ws}. Note however, that many such solutions suffer from perturbative instabilities \cite{Donos:2011pn,NPB2011} which sheds some doubt on their physical relevance.

\item The holographic description of our supergravity solutions, when embedded in string or M-theory, should be in terms of a partially topologically twisted SCFT on a Riemann surface. Understanding how the RG flow across dimensions in the SCFT realizes the uniformization of the Riemann surface metric is presumably a hard question in the strongly coupled QFT description, see for example \cite{Gaiotto:2011xs}. 

\item The four-dimensional black hole solutions we studied have a near-horizon AdS$_2$ geometry which is universal and independent of the Riemann surface metric away from the horizon. The metric deformations of $\Sigma_{\mathfrak{g}}$ can be interpreted as irrelevant operators in the one-dimensional IR theory holographically dual to this near-horizon region which determine the details of the deformations performed in the UV three-dimensional SCFT. It will be interesting to understand whether our solutions have a relation to the recent studies of near AdS$_2$ holography, see for example \cite{Almheiri:2014cka,Maldacena:2016upp}.

\end{itemize}

\bigskip
\bigskip
\leftline{\bf Acknowledgements}
\smallskip
\noindent We are grateful to Mike Anderson, Chris Beem, Francesco Benini, Pieter Bomans, Davide Cassani, Anthony Charles, Marcos Crichigno, Kiril Hristov, Vincent Min, Krzysztof Pilch, Leonardo Rastelli, Val Reys, and Balt van Rees for interesting discussions. The work of NB is supported in part by an Odysseus grant G0F9516N from the FWO. FFG is supported by a postdoctoral fellowship from the Fund for Scientific Research - Flanders (FWO). KP is supported in part by a doctoral fellowship from the Belgian American Educational Foundation. NB and FFG are also supported by the KU Leuven C1 grant ZKD1118 C16/16/005.

 \newpage
 \appendix
 \section{Universal BPS equations}
 \label{UniversalDerivation}
 
In this Appendix we present some details on the derivation of the BPS equations in the minimal four-, five-, and six-dimensional gauged supergravity theories used for the analysis in Section~\ref{sec:universal}. The derivation of the BPS equations for the seven-dimensional minimal gauged supergravity was presented in \cite{Bah:2012dg}. We also note that we have explicitly confirmed that the equations of motion for the Ansatz we study are implied by the BPS equations.
 
\subsection{Four dimensions}
\label{4Dapp}
The Lagrangian for the bosonic fields of the minimal $\mathcal{N}=2$ four-dimensional gauged supergravity is
\begin{equation}\label{eq:4dminLagapp}
    \mathcal{L} = R + 6  - \frac34 F_{\mu\nu}F^{\mu\nu}\,.
\end{equation}
The gravitino variations are given by 
\begin{align}
    \delta \psi^i_\mu = \nabla_\mu \epsilon^i - A_\mu \varepsilon^{ij}\epsilon^j+ \frac12 \gamma_\mu \epsilon^i + \frac14F_{\nu\lambda}\gamma^{\nu \lambda} \gamma_\mu \varepsilon^{ij}\epsilon^j\,.
\end{align}
We consider the following Ansatz
\begin{equation}\label{eq:4dAnsatzapp}
\begin{split}
\rmd{s}^2_4 =&~ -\rme^{2f(r,x,y)}\rmd{t}^2 + \rme^{2h(r,x,y)}\rmd{r}^2+ {\rme^{2g_1(r,x,y)}}\rmd{x}^2+{\rme^{2g_2(r,x,y)}}\rmd{y}^2\,,\\
A =&~ A_x(r,x,y) \rmd{x} + A_y(r,x,y) \rmd{y}\,. 
\end{split}
\end{equation}
Notice that this is the most general static Ansatz assuming an isometry in the $t$-direction. We are interested in solutions of this supergravity theory which preserve $1/4$ of the supersymmetry. This leads to the following projectors
\be
\gamma_{\Hat{r}} \epsilon^i = \epsilon\,\quad\text{and}\quad \gamma_{\Hat{x}\Hat{y}}\epsilon^i = -\varepsilon^{ij}\epsilon^j\,,
\ee
where we use a hat to denote tangent space indices. Notice that we allow for the spinor parameter, $\epsilon^i$, to be depend on the coordinates $(r,x,y)$. The BPS equations can be derived by considering a linear combination of the supersymmetry variations for which no derivative of the spinor parameter appears. In particular the combination
\be
\gamma_{\Hat{x}\Hat{y}}\delta \psi^i -\varepsilon^{ij}\delta \psi^j~,
\ee
leads to the equations
\be\label{4dmin: con1}
\begin{split}
\e^{h+g_1+g_2}(\partial_x + i\partial_y)f &= -\e^{2g_2}F_{rx}+i \e^{2g_2}F_{ry}\,,\\
(\partial_x + i\partial_y) (f+h)&=0\,,\\
\e^{g_1+g_2}\partial_r g_1 &= \e^{h}(\e^{g_1+g_2}-F_{xy})\,,\\
\partial_r (g_1-g_2) &= 0\,,
\end{split}
\ee
where $F_{rx}$, $F_{ry}$, and $F_{xy}$ are the non-trivial component of the gauge field 2-form flux $F=dA$. The $t$-direction of the gravitino variation leads to one more condition:
\be
\partial_r(f+g_1) =-2\e^h,\label{4dmin: con2}\,.
\ee
Now it is easy to read off from the gravitino variation  differential equations for the spinor parameter itself
\be
\partial_{r,x,y} \epsilon^i = \f12 (\partial_{r,x,y} f)\epsilon^i\,,
\ee
as well as an equation for the gauge field 
\be\label{4dmin: gauge}
A_x = -\frac12 \e^{g_1-g_2}\partial_y (f+g_1)\,, \qquad A_y = \frac12 \e^{g_2-g_1}\partial_x (f+g_2)\,.
\ee

We can now solve these equations and show that they imply \eqref{4Dsolution} and \eqref{4Dmaster}. First we notice that
\be\label{4Dgfunc}
g_2 = g_1 + C(x,y)\,,
\ee
for some function $C$ on the Riemann surface. We can always choose coordinates on the Riemann surface such that $C=0$. This in turn implies 
\be\label{4Dgfuncsol}
g_1 = g_2 \equiv g\,.
\ee
This considerably simplifies the remaining equations. In particular, since $f+h$ is a function only of $r$ we are free to change to a new radial coordinate $\rho(r)$ defined by 
\be
\f{\dd \rho}{\dd r} \equiv-\rme^{f+h}~,
\ee
In these new coordinates we obtain the following two equations by using $F_{xy} = \partial_x A_y - \partial_y A_x$ and \eqref{4dmin: gauge} in addition to \eqref{4dmin: con1}
\begin{equation}\label{4dmin: 2eqns}
    \begin{split}
    2\rme^{-f} - \partial_\rho(f+g) =0\,,\\
    2(-\rme^{f}\partial_\rho f +1) + \rme^{-2g}(\partial^2_x + \partial^2_y)(f+g) =0\,.
    \end{split}
\end{equation}
We also define the new function $2\varphi = f+g$. The first equation above then determines $f$ as a function of $\varphi$ which allows us to write the metric and gauge field purely in terms of $\varphi$ as in \eqref{4Dsolution}. The second equation reduces to a single PDE for $\varphi$:
\begin{equation}\label{4dmin: nice}
    \boxed{~~\triangle \varphi +\e^{4\varphi}\Big(\partial_\rho^2 \varphi + (\partial_\rho \varphi)^2\Big)=0\,,~~}
\end{equation}
where $\triangle = \partial^2_x + \partial^2_y$.
This is the flow equation we use in Section \ref{4DUniversal}. Finally we note that the supersymmetry spinor parameter can be found explicitly and takes the form 
\be \label{eq:spinoref4d}
\epsilon^i = \e^{f/2}\epsilon^i_0\,,
\ee
where $\epsilon^i_0$ is a constant spinor obeying the projectors
\be 
\gamma_{\Hat{r}} \epsilon_0^i = \epsilon_0^i\,\quad\text{and}\quad \gamma_{\Hat{x}\Hat{y}}\epsilon^i_0 = -\varepsilon^{ij}\epsilon^j_0\,.
\ee
Note that the form of the spinor in \eqref{eq:spinoref4d} is compatible with the fact that the spinor bilinear is proportional the time-like Killing vector $\partial_{t}$ in \eqref{eq:4dAnsatzapp}.

Before we move on to the presentation of the derivation of the BPS equations in the five- and six-dimensional supergravity theories we emphasize two important results. First, we have shown that for both of these theories the functional dependence of the spinor on $(r,x,y)$ is fixed entirely in terms of the function $f$ as in \eqref{eq:spinoref4d}.  Similarly we have shown that the metric functions $g_1$ and $g_2$ are always related as in \eqref{4Dgfunc}. To simplify the analysis and avoid repetition we will present the derivation of the BPS equations in five and six dimensions with \eqref{eq:spinoref4d} and \eqref{4Dgfunc} implemented from the start. 

\subsection{Five dimensions}
\label{5Dapp}
The bosonic part of the Lagrangian for the five-dimensional minimal gauged supergravity is given by 
\begin{equation}
\mathcal{L} = R + 12 - \frac34 F_{\mu\nu}F^{\mu\nu} + \frac14 \epsilon^{\mu\nu\alpha\beta\rho}F_{\mu\nu}F_{\alpha\beta}A_\rho\,,
\end{equation} 
and the gravitino variations are given by
\begin{equation}
    \delta \psi_\mu = \left(\nabla_\mu + \frac{i}{8}(\gamma^{\nu\rho}_\mu - 4 \delta^\nu_\mu \gamma^\rho)F_{\nu\rho}+\frac12 \gamma_\mu -\frac{3i}{2}A_\mu\right)\epsilon\,.
\end{equation}
We consider the following Ansatz 
\be
\begin{split}
\rmd{s}^2_5 =&~ \rme^{2f(r,x,y)}\dd s_{\mathbf{R}^{1,1}}^2 + \rme^{2h(r,x,y)} \rmd{r}^2+ \rme^{2g(r,x,y)}(\rmd{x}^2+\rmd{y}^2)\,,\\
A =&~ A_x(r,x,y) \rmd{x} + A_y(r,x,y) \rmd{y}\,,
\end{split}
\ee
where $\dd s_{\mathbf{R}^{1,1}}^2$ is the metric on the two-dimensional Minkowski space. Furthermore, we impose the following projectors for the supersymmetry spinor parameter $ \epsilon = \rme^{f/2}\epsilon_0$
\begin{equation}
  \gamma_{\hat{r}}\epsilon_0 = \epsilon_0, \hspace{0.5cm} \gamma_{\hat{x}\hat{y}}\epsilon_0 = -i\epsilon_0\,.
\end{equation}
The $t$-component of the gravitino variation the yields
\begin{equation}\label{5dmin: bpst1}\begin{split}
    2\rme^{-h}\partial_r f + 2 +  \rme^{-2g}F_{xy} &= 0\,,\\
2 \rme^{h}(\partial_x + i \partial_y)f + i(F_{rx} + iF_{ry}) &=0\,.
\end{split}\end{equation}
This determines the field strength in terms of the geometry. The other BPS constraints, after having substituted the field strength wherever it appears, lead to the constraints
\begin{eqnarray}\label{5dmin: con1}\begin{split}
(\partial_x + i\partial_y)(2f+h) &=& 0\,,\\
3\rme^h + \partial_r (2f + g) &=&0\,,
\end{split}    
\end{eqnarray}
together the following expression for the gauge fields
\begin{equation}\label{5dmin: gauge}
A_x = -\frac13 \partial_y(2f+g)\,, \hspace{0.5cm}A_y = \frac13\partial_y (2f+g)\,.
\end{equation}

Since $2f+h$ is a function only of the coordinate $r$ we are free to choose a new radial variable $\rho(r)$ which is defined by $\rmd{\rho} = -\rme^{2f+h} \rmd{r}$. Using this we obtain the system of equations
\begin{equation}\label{5dmin: 2eqns}
    \begin{split}
     6(-\rme^{2f}\partial_\rho f + 1)+ \rme^{-2g}(\partial^2_x + \partial^2_y)(2f+g) = 0\,,\\
     3\rme^{-2f} = \partial_\rho(2f+g)\,.
    \end{split}
\end{equation}
The latter equation can be used to replace $f$ by the new variable $3\varphi = 2f + g$ which due to the first equation in \eqref{5dmin: 2eqns} satisfies
\begin{equation}\label{5dmin: nice}
   \boxed{~~\triangle \varphi +\e^{6\varphi}\Big(\partial_\rho^2 \varphi + 2(\partial_\rho \varphi)^2\Big) = 0\,.~~}
\end{equation}
This is the PDE used in the analysis in Section~\ref{subsec:5dmin}.

\subsection{Six dimensions}
\label{6Dapp}

We work with the six-dimensional $\mathcal{N}=4$ $\SU(2)$ gauged supergravity constructed in \cite{Romans:1985tw}. The bosonic sector of the theory contains a graviton $e^a_\mu$, three $\SU(2)$ gauge fields $A^I_\mu$, one Abelian gauge field $a_\mu$, a two-index tensor gauge field $B_{\mu\nu}$ and a real scalar field $\alpha$. For the solutions of interest here we can consistently take the Abelian gauge field $a_\mu$ and the tensor field $B_{\mu\nu}$ to vanish. In addition we only need to turn on the Cartan generator of the $\SU(2)$ gauge field, i.e. take only $A^3_\mu$ to be non-zero. We note that there are two coupling constants, $g$ and $m$, in the construction of  \cite{Romans:1985tw}. We are interested in the so called $\mathcal{N}=4^+$ case, where $g>0$ and $m>0$ since this theory has a supersymmetric AdS$_6$ vacuum solution with $g = 3m$. We fix the length scale of AdS$_6$ by setting $m=\sqrt{2}$. The resulting bosonic Lagrangian is then given by\footnote{Note that we use similar conventions to \cite{Naka:2002jz} and work in mostly plus signature.} 
\begin{equation}
   \mathcal{L} = R - 4(\partial_\mu \alpha)^2 - 2\rme^{-2\alpha}F^{\mu\nu}F_{\mu\nu} + (9 \rme^{2\alpha} + 12 \rme^{-2\alpha}-\rme^{-6 \alpha})\,. 
\end{equation}
The supersymmetry transformations for the fermionic fields, with $i=1,2$, are
\begin{equation}
\begin{split}
    \delta \psi_{\mu i}& = \nabla_\mu \epsilon_i -3i A_\mu \epsilon_i +\frac18 (3\rme^\alpha + \rme^{-3\alpha})\gamma_\mu \gamma_7 \epsilon_i+\frac{i}{8}\rme^{-\alpha}(\gamma^{\nu\rho}_\mu-6\delta^\mu_\nu \gamma^\rho)F_{\mu\nu}\epsilon_i\,, \\
    \delta \chi_i &= \gamma^\mu \partial_\mu \alpha \epsilon_1 - \frac34(\rme^\alpha - \rme^{-3\alpha})\gamma_7 \epsilon - \frac{i}{4}\rme^{-\alpha}\gamma^{\mu\nu}F_{\mu\nu}\gamma_7 \epsilon_i\,,
    \end{split}
\end{equation}
Here we consider the following Ansatz 
\begin{equation}
   \begin{split}
\rmd{s}^2 &= \rme^{2f(r,x,y)}\dd s_{\mathbf{R}^{1,2}}^2 +\rme^{2h(r,x,y)}\rmd{r}^2 +\rme^{2g(r,x,y)}(\rmd{x}^2 + \rmd{y}^2)\,,\\
A &= A_x(r,x,y) \rmd{x} + A_y(r,x,y) \rmd{y}\,,\\
\alpha &= \alpha(r,x,y)\,,
\end{split} 
\end{equation}
where $\dd s_{\mathbf{R}^{1,2}}^2$ is the metric on the three-dimensional Minkowski space. Using this Ansatz in the supersymmetry variations we obtain the BPS equations by taking $\epsilon = \rme^{f/2}\epsilon_0$ and imposing the projectors $\gamma_{\hat{r}} \epsilon_0 = \gamma_7 \epsilon_0$ and $\gamma_{\hat{x}\hat{y}}\epsilon_0 = i\epsilon_0$.  The spin-$1/2$ variations yield
\begin{eqnarray}\label{6dmin: dil1}
    (\partial_x - i\partial_y)\alpha + \frac{i}{2}\rme^{-\alpha-h}(F_{rx}-iF_{ry})&=&0\,,\\
\label{6dmin: dil2}
 \partial_r \alpha - \frac34 \rme^h(e^\alpha - \rme^{-3\alpha})-\frac12 \rme^{h-\alpha-2g}F_{xy}&=&0\,.   
\end{eqnarray}
We shall use these relations to replace the field strength wherever it appears in the spin-$3/2$ variations. The variations with components along $\dd s_{\mathbf{R}^{1,2}}^2$ result in
\begin{equation}\label{6dmin: gravt1}
(\partial_x +i\partial_y)(f+\alpha)=0, \hspace{0.5cm} \partial_r(f+\alpha) + \rme^{h-3\alpha}=0\,.
\end{equation}
This in turn implies that $f+\alpha$ and $h-3\alpha$ do not depend on $x,y$. The other components of the spin-$3/2$ variation then yield the constraint 
\begin{equation}\label{6dmin: con1}
    -2 \rme^{h-3\alpha}+3\rme^{h+\alpha}+\partial_r(g-3\alpha) = 0\,,
\end{equation}
together with the following expression for the gauge fields
\begin{equation}
    A_x = -\frac16 \partial_y (g-3\alpha)\,, \hspace{0.5cm} A_y = \frac16 \partial_x (g-3\alpha)\,.
\end{equation}
As is familiar by now we can change the radial coordinate by using $\rmd{r} = - \rme^{3\alpha-h}\rmd{\rho}$ to obtain the pair of equations:
\begin{equation}\label{6dmin: 2eqns}
    \begin{split}
    12 \rme^{-3\alpha} \partial_\rho \alpha +9(\rme^\alpha - \rme^{-3\alpha}) + \rme^{-\alpha-2g}(\partial^2_x + \partial^2_y)(g-3\alpha) = 0\,,\\
    \partial_\rho (g-3\alpha) + 2 - 3\rme^{4\alpha}=0\,.
    \end{split}
\end{equation}
We can define $3\varphi = g - 3\alpha + 2\rho$ and use the second equation in \eqref{6dmin: 2eqns} to find
\begin{equation}
    \partial_\rho \varphi =  \rme^{4\alpha}\,.
\end{equation}
This relations determines uniquely $\alpha$ from $\varphi$. Moreover, it enables us to rewrite the first equation of \eqref{6dmin: 2eqns} in terms of $\varphi$ alone, 
\begin{equation}\label{6dmin: nice}
    \boxed{~~\triangle \varphi +\e^{6\varphi}\e^{-4\rho}\big(\partial_\rho^2 \varphi + 3 (\partial_\rho \varphi)^2 - 3 \partial_\rho \varphi \big)=0\,.~~}
\end{equation}
This is the PDE used in the analysis in Section~\ref{subsec:6dmin}.

\newpage
\bibliography{refs}
\bibliographystyle{JHEP}

\end{document}